\documentclass[preprint]{aastex}
\usepackage{latexsym,bm}
\usepackage{aecompl}
\usepackage{graphicx}   
\usepackage{amsmath}    
\usepackage{amssymb}    
\usepackage{subfigure}
\usepackage{booktabs}

\begin{document}

\title{AE Ursae Majoris -- a $\delta$ Scuti Star in the Hertzsprung Gap}
\author{Jia-Shu Niu $^{1,2,3}$, 
Jian-Ning Fu $^{1}$\thanks{E-mail: jnfu@bnu.edu.cn}, 
Yan Li $^{4}$, 
Xiao-Hu Yang $^{1,5}$,
Weikai Zong$^{1,6,7}$,
Hui-Fang Xue $^{1}$,
Yan-Ping Zhang $^{1}$,
Nian Liu $^{1}$,
Bing Du $^{5}$
and Fang Zuo$^{5}$.}
\affil{$^{1}$Astronomy Department, Beijing Normal University,    Beijing 100875, P.R.China\\
$^{2}$CAS Key Laboratory of Theoretical Physics and Kavli Institute for Theoretical Physics China (KITPC), Institute of Theoretical Physics, \\ Chinese Academy of Sciences, Beijing, 100190, P.R.China\\
$^{3}$School of Physical Sciences, University of Chinese Academy of Sciences, No.19A Yuquan Road, Beijing 100049, P.R.China\\
$^{4}$National Astronomical Observatories/Yunnan Observatory, Chinese Academy of Sciences, PO Box 110, Kunming 650011, P.R.China\\
$^{5}$National Astronomical Observatories, Chinese Academy of Sciences, Beijing 100012, P.R.China\\
$^{6}$Universit\'e de Toulouse, UPS-OMP, IRAP, Toulouse, France\\
$^{7}$CNRS, IRAP, 14 avenue Edouard Belin, 31400 Toulouse, France
}

\begin{abstract}
We analyze the photometric data and spectroscopic data that collect on the $\delta$ Scuti star AE UMa. The fundamental and the first overtone frequencies are confirmed as $f_{0} = 11.62560$\,c\,d$^{-1}$ and $f_{1} = 15.03124$\,c\,d$^{-1}$, respectively, from the frequency content by analyzing of the 40 nights light curve spanning from 2009 to 2012. Additionally, another 37 frequencies are identified as either the harmonics or the linear combinations of the fundamental and the first overtone frequencies, among which 25 are newly detected.
 The rate of period change of the fundamental mode is determined as $(1/P_{0})(dP_{0}/dt) = 5.4(\pm 1.9) \times 10^{-9}$\,yr$^{-1}$ as revealed from the $O-C$ diagram based on the 84 newly determined times of maximum light combined with those derived from the literature. The spectroscopic data suggests that AE UMa is a population\,I $\delta$ Scuti star. With these physical properties, we perform theoretical explorations based on the stellar evolution code MESA on this target, considering that the variation of pulsation period is caused by secular evolutionary effects. We finally constraint the AE UMa with the physical parameters as: the mass of $1.805 \pm 0.055\ M_{\odot}$, the radius of $1.647 \pm 0.032 \times 10^{11}$\,cm, the luminosity of $1.381 \pm 0.048~(\log L/L_{\odot})$ and the age of $1.055 \pm 0.095 \times 10^{9}$\,yr. AE UMa can be the (Pop. I) $\delta$ Scuti star that locates just after the second turn-off of its evolutional track leaving the main sequence, a star in the phase of the Hertzsprung Gap with a helium core and a hydrogen-burning shell.
\end{abstract}


\section{INTRODUCTION}

$\delta$ Scuti stars are a class of pulsating variable stars that lie in the classical instability strip crossing the main sequence on the Hertzsprung-Russell diagram. Their pulsations are driven by the $\kappa$-mechanism which drives both the Cepheids and the RR Lyrae stars as well. The amplitudes of pulsations in $\delta$\,Scuti stars are from mmag up to tenths of a magnitude,  periods between $0.03$ and $0.3$\,days \citep[see, e.g.,\ ][]{Niu2013,Zong2015}.  These stars are found with masses between $1.5$ and $2.5\ M_{\odot}$, luminosities between $10$ and $50\ L_{\odot}$.  The general consensus show that most (possibly all) $\delta$ Scuti stars are normal stars which evolve in the main-sequence or the immediate post-main-sequence stages, according to standard stellar-evolution theory \citep[see, e.g.,\ ][]{Baglin1973,Breger1979,Breger1980}. Nevertheless, observational proof of the validity of this hypothesis has not been found yet \citep{Petersen1996}.\\

The high-amplitude $\delta$ Scuti stars (hereafter HADS) are traditionally found with slow rotation, one or two dominant radial modes with amplitudes larger than 0.1 mag, although some of them may have low-amplitude nonradial modes \citep[e.g.,][]{Poretti2003}.
SX Phoenicis (SX Phe) stars is a subgroup of HADS with low metallicity and large spatial motion \citep[see e.g.,][]{Fu200800}. They are old Pop. II stars and found to be members of globular clusters \citep{Rodriguez200008}. However, some of them have been discovered in the general star fields \citep{Rodriguez2001}. Interestingly, pulsations in the majority of the field SX Phe variables display simple frequency spectra with short periods ($\leq0^{\rm d}.08$) and large visual peak-to-peak amplitudes \citep[$\geq0^{\rm m}.1$, e.g.,][]{Fu2008}. The period changes of pulsations can be determined based over long-term and high-precision photometric observations on such stars, which can constraint the stellar evolutionary phase of the star \citep[e.g.,][]{Yang2012}.\\

The star AE Ursae Majoris (hereafter AE UMa = HIP 47181, $\alpha_{2000} = 09^{\rm h}36^{\rm m}53^{s}$, $\delta_{2000} = 44^{\circ}04'01''$, $<V> = 11^{\rm m}.27$, $P_{0} = 0^{\rm d}.0860$, $\Delta V = 0^{\rm m}.10$), was discovered to be a variable star by \citet{Geyer1955}. The spectral type of AE UMa was classified in accordance with the type of variability by \citet{Gotz1961} as A9.  The period of light variations was determined by \citet{Tsesevich1973} and they classified it as a dwarf Cepheid. The beat phenomenon of the pulsations of this star was found by \citet{Szeidl1974}. AE UMa was listed as an SX Phe star by \citet{Garcia1995}. However, \citet{Hintz1997} showed strong evidence against this classification and reclassified it as a normal Population I, high-amplitude $\delta$ Scuti star. According to the measurement of \citet{Breger1998}, AE UMa had fast period decreasing rate of $4.8 \times 10^{-7}\ {\rm yr}^{-1}$ hence it should be a pre-MS star. However, there is no other evidence for this star to be a pre-MS star. Recently, both \citet{Pocs2001} and \citet{Zhou2001} analyzed pulsations of the star with high-precision and longer photometric data. Their results are consistent with the classification with the outcomes of \citet{Hintz1997}: AE UMa is a Pop. I, post-MS $\delta$ Scuti star, but with a stable fundamental frequency and the first overtone decreasing with a rate of $\sim 10^{-8}\ {\rm yr}^{-1}$.\\

In this paper, we present a detailed study of the pulsations and the period changes of AE UMa, mainly based on both photometric observations and spectroscopic observations. Based on the observational results, we perform theoretical explorations using the stellar code MESA and constraint the physical parameters on this star. The organization of the paper is: Section 2 describes  photometry and data reduction,  as well as spectral results; we present the pulsation analysis of the new data in Section 3; in Section 4, the rate of period change of the fundamental pulsations is determined before we conduct calculations of the stellar models with the constraints of the stellar parameters, the frequencies and their variations in Section 5; The conclusions of the study is given in the final section.

\section{OBSERVATIONS AND DATA REDUCTION}
Photometric observations for AE UMa were made with the 85-cm telescope located at the Xinglong Station of NAOC between March 2009 and May 2012. The 85-cm telescope was equipped with a standard Johnson-Cousin-Bessel multicolour filter system and a PI1024 BFT CCD camera mounted on the primary focus \citep{Zhou2009}. The CCD had $1024 \times 1024$ pixels, corresponding to a field of view of $16.5^{'} \times 16.5^{'}$. Since March 2012, the CCD camera has been replaced by a PI512 BFT with larger size of pixels, which has $512\times512$ pixels corresponding to a field of view of $15^{'} \times 15^{'}$. The observations were made through a standard Johnson $V$ filter with the exposure time ranging from 15 to 120 seconds, depending on the atmospheric conditions. A journal of the new observations is listed in Table \ref{tab1}.

\begin{table*}
\caption{Journal of photometric observations in $V$ for AE UMa with the 85-cm telescope.}
\centering
\centerline{}
 \begin{tabular}{ccccc}
 \hline
  CCD & Year & Month & Nights & Frames \\

 \hline
   PI BFT1024 & 2009 & Mar & 5 & 4055    \\
              & 2009 & May & 3 & 516   \\
              & 2010 & Feb & 2 & 1385    \\
              & 2011 & Jan & 5 & 1275   \\
              & 2011 & Feb & 8 & 4759   \\
              & 2012 & Jan & 1 & 328   \\
              & 2012 & Feb & 6 & 1887    \\
 \hline
   PI BFT512  & 2012 & Mar & 5 & 2516   \\
              & 2012 & Apr & 5 & 556   \\
 \hline
 \end{tabular}
\label{tab1}
 \end{table*}

\begin{figure*}
\centering
\includegraphics[width=0.6\textwidth,height=0.6\textwidth]{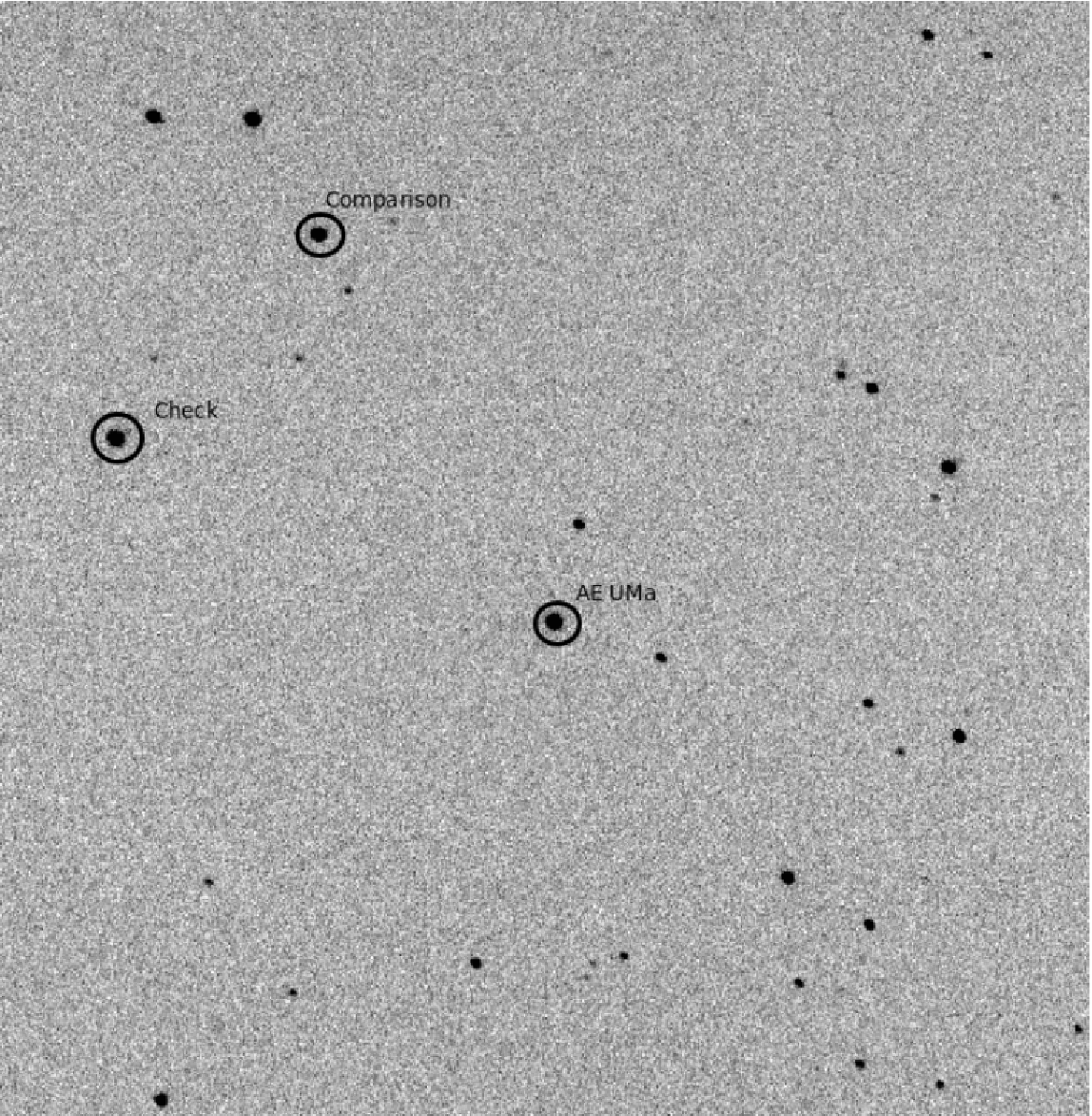}
\caption{A CCD image ($16.5^{'}\times 16.5^{'}$) of AE UMa ($\alpha_{2000} = 09^{\rm h}36^{\rm m}53^{\rm s}$, $\delta_{2000} = +44^{\circ}04^{'}00^{''}$) taken with the 85-cm telescope at the Xinglong Station. North is down and East is to the right. AE UMa, the comparison and the check star are marked.}
\label{fig1}
\end{figure*}

In total, 17277 CCD frames were collected for AE UMa during 40 nights. Figure \ref{fig1}  shows an image of AE UMa taken with the 85-cm telescope, where the comparison star (TYC 2998-1249-1) and the check star (TYC 2998-1166-1) are marked as well. The details of the three stars from SIMBAD \citep{Wenger2000} are listed in Table \ref{tab2}.

\begin{table*}
\caption{The comparison star and the check star used in the photometry of AE UMa}
\centering
\scalebox{0.8}[0.8]
{
\begin{tabular}{cccccc}
\hline
Star name    &$\alpha$(2000)    &$\delta$(2000)    &$V$    &$B$   &$B-V$\\
\hline
\hline
Object = AE Ursae Majoris & $09^{h}36^{m}53^{s}.155$ & $+44^{\circ}04^{'}00^{''}.39$ & $11.35 \pm 0.08$ & $11.54 \pm 0.06$ & $0.19 \pm 0.14$\\
Comparison = TYC 2998-1249-1  &$09^{h}37^{m}28^{s}.5826$  &$+44^{\circ}01^{'}16^{''}.854$  & $11.32 \pm 0.08$ & $11.82 \pm 0.08$ & $0.50 \pm 0.16$  \\
Check = TYC 2998-1166-1  &$09^{h}37^{m}12^{s}.058$  &$+43^{\circ}58^{'}20^{''}.12$  & $12.21 \pm 0.18$ & $13.10 \pm 0.30$ & $0.89 \pm 0.48$  \\
\hline
\end{tabular}
}
\label{tab2}
\end{table*}

The preliminary processing of the CCD frames (bias, dark substraction and flat field correction) was performed with the standard routines of CCDPROC from the IRAF software. After that, we employed the IRAF DAOPHOT package to perform aperture photometry. In order to optimize the size of the aperture, we used 12 different size of apertures for the data in each night and adopted the aperture which brought the minimum variance of the magnitude differences between the check star and the comparison star. The data reduction was carried out with the standard process of aperture photometry.

The light curves were then produced by computing the magnitude differences between AE UMa and the comparison star. The standard deviations of the magnitude differences between the check star and the comparison star yielded an estimation of photometry precisions, with the typical value of $0^{\rm m}.003$ in good observation conditions and $0^{\rm m}.011$ in poor cases from night to night. As there were slight zero-point shifts, we adjusted it with the fitted light curves for every month by assuming that the pulsations were stable in one month.

\begin{figure*}
\begin{center}
\includegraphics[width=1.0\textwidth,height=0.8\textheight]{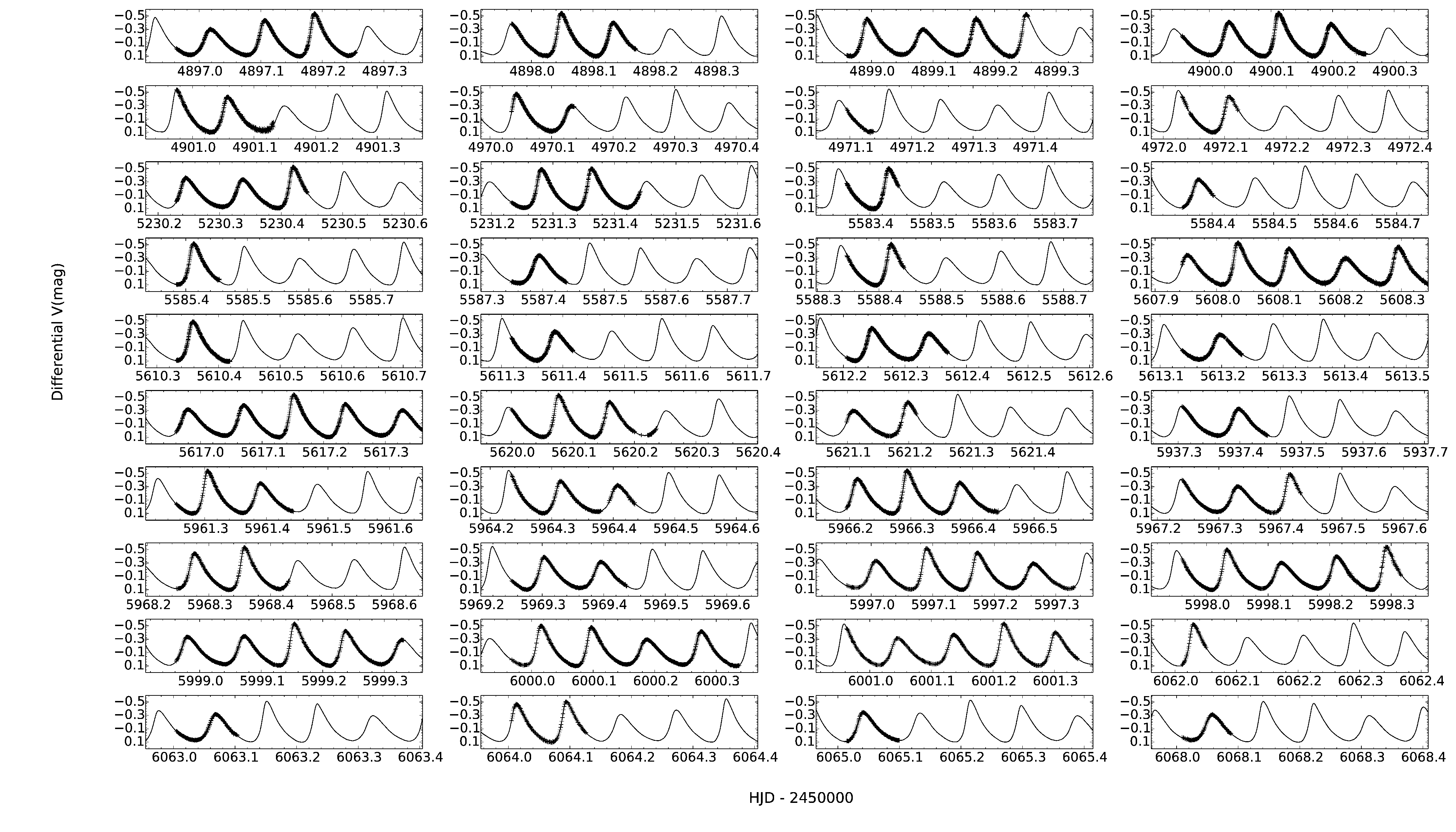}
\end{center}
\caption{Light curves of AE UMa relative to the comparison star in the $V$ band from 2009 to  2012, observed with the 85-cm telescope. The solid curves represent the fitting with a solution up to 18 frequencies listed in Table \ref{tab3}.}
\label{fig2}
\end{figure*}

Figure \ref{fig2} shows the light curves of AE UMa in Johnson $V$ band observed with the 85-cm telescope in 2009, 2010, 2011 and 2012, which was used to make pulsation analysis, and determination of new times of maximum light.

Spectroscopic observation for AE UMa was made with the 2.16 m telescope which locates at Xinglong station of NAOC on May 21, 2016. The BFOSC low-dispersion spectrometer was employed from the observations. The used grating was G7 with a slit width of $1.8^{''}$ and a line dispersion of 95 ${\mathring A}$/mm. The center wavelength was at 530 nm with the wavelength range of 380-680 nm.

\begin{figure*}
\centering
\includegraphics[width=0.8\textwidth,height=0.4\textheight]{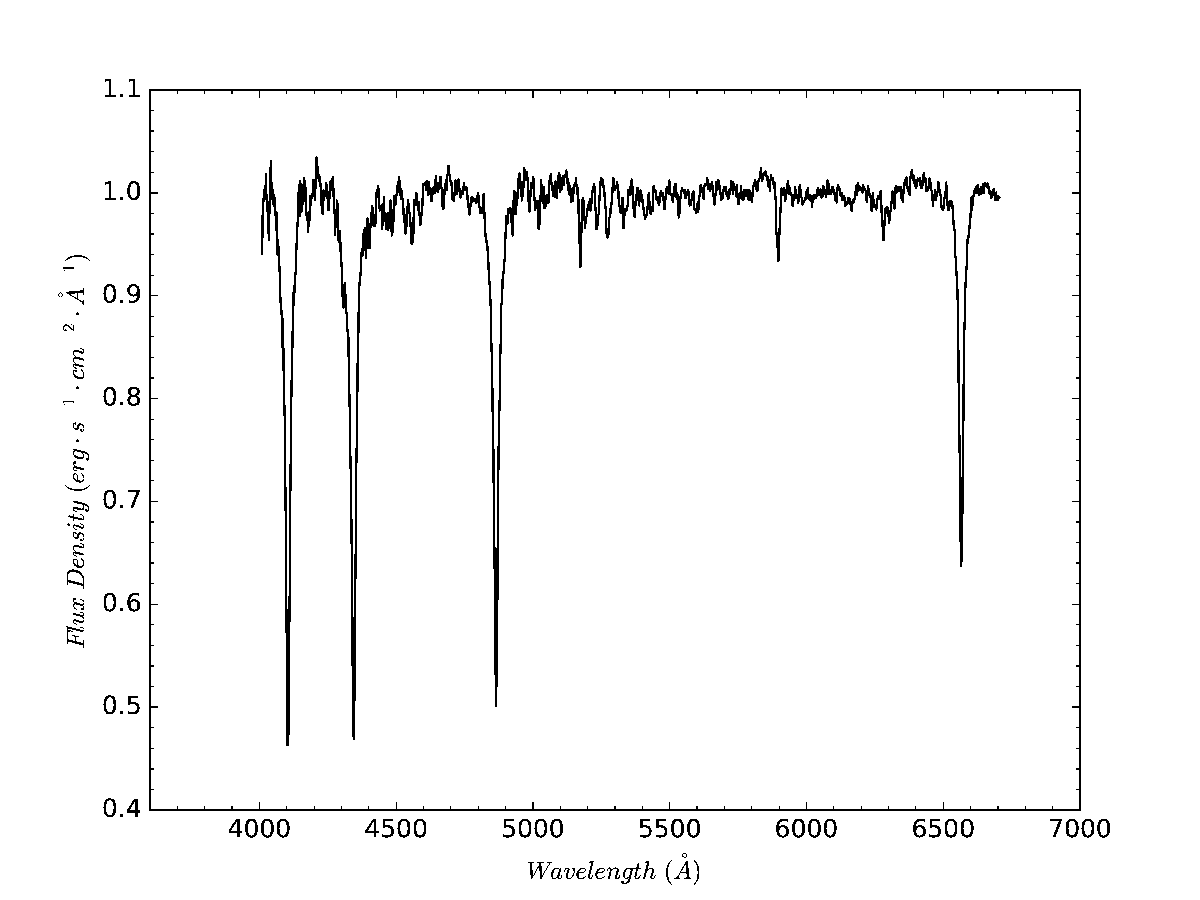}

\caption{Spectroscopic observations results of AE UMa.}
\label{fig3}
\end{figure*}

The data were reduced with IRAF and the obtained low-resolution spectrum is shown in Figure \ref{fig3}. With the results from spectroscopic observation, we used the automated 1D parametrization pipeline LASP which base on the stellar spectral template library \citep{Wu2011} to get the stellar atmospheric parameters (see Table \ref{tab3}).

\begin{table*}
\caption{Parameters of AE UMa derived from spectroscopic observations.}
  \centering 
\begin{tabular}{|c|c|c|}
 \hline
  Parameters  & Values  & $\sigma$ \\

 \hline
  $T_\mathrm{eff}\ (K)$ & 7600 & 180 \\
  $\log g$  & 4.1 & 0.2  \\
  $[Fe/H]$  & -0.32  & 0.23 \\
  $RV\ (km/s)$  & 150 & 27 \\
\hline
 \end{tabular}
\label{tab3}
\end{table*}

 We note that the $T_{\rm eff}$ and $\log g$ correspond to a fixed phase since we only had acquired one spectrum.
 These values may vary differently from phase to phase during the pulsations of AE UMa.
 However, the metal to hydrogen ratio $[Fe/H]$ is not sensitive to the pulsations.  The value of $[Fe/H]$ is $-0.32 (\pm 0.23)$ which indicate that AE UMa is possibly a Pop. I $\delta$ Scuti star. This is consistent with the classification of the results obtained from \citet{Hintz1997}, \citet{Pocs2001} and \citet{Zhou2001} without spectroscopic data. Hence, AE UMa can be modeled as a single star in Section 5.

\section{PULSATION ANALYSIS}
Pulsation analysis was performed with the light curves of AE UMa in the years 2009, 2010, 2011 and 2012, respectively with the software PERIOD04 \citep{Lenz2005}, which provides Fourier transformations of the light curves to search for significant peaks in the amplitude spectra until 150 c\,d$^{-1}$, since there is no significant ones above this frequency limit. Then, the light curves are fitted with the following formula,
\begin{equation}
m=m_{0}+ \Sigma A_{i}\sin(2\pi(f_{i}t+\phi_{i})).
\end{equation}

Table~\ref{tab4} lists the solutions of 37 frequencies whose signal-to-noise ratios (S/N) are higher than 4.0 \citep{Breger1993} and the averaged noise level is calculated over the whole frequency range, 0-150~c\,d$^{-1}$ \citep[e.g.,][]{Kepler2005}. The solid curves in Figure \ref{fig2} show the fits with the frequency solutions in different years. From Table \ref{tab4}, one notes that the 37 frequencies are composed of the fundamental and the first overtone frequencies, their harmonics and linear combinations. As can be noticed, no significant signals are detected in addition to these frequencies.

\begin{table*}
\caption{Multi-frequency Solutions of the Light Curves of AE UMa in $V$ Band in 2009, 2010, 2011 and 2012. Fre: Frequency in $c\ d^{-1}$. Amp: Amplitude in $mmag$. S/N: signal to noise ratio.}
\centering
  \scalebox{0.65}[0.6]{
 \begin{tabular}{c|c|ccc|ccc|ccc|ccc}
\hline
\hline
NO.&\textbf{Marks} & Fre & Amp & S/N & Fre & Amp & S/N & Fre & Amp & S/N & Fre & Amp & S/N\\
\hline
& \multicolumn{1}{c}{ }\vline& \multicolumn{3}{c}{2009} \vline& \multicolumn{3}{c}{2010} \vline& \multicolumn{3}{c}{2011} \vline& \multicolumn{3}{c}{2012}\\

\hline

     1& $f_{0}$&11.62525&       220.45& 1078.98&  11.62944&       216.36& 2343.67& 11.62549&       218.93& 1360.27& 11.62558&       218.26& 1652.90\\
     2& $2f_{0}$&23.25014&      74.36&  373.67&  23.25888&       73.25&  727.33& 23.25036&       73.77&  467.10&  23.25108&       73.57&  565.07\\
     3& $f_{1}$&15.03109&       45.29&  223.73&  15.01105&       43.95&  464.85&  15.03201&       45.00&  278.92&  15.03117&       45.11&  343.71\\
     4& $3f_{0}$&34.87873&      28.48&  147.81&   34.88832&       29.66&  299.18& 34.87620&       27.70&  175.13&  34.87680&       27.99&  216.03\\
     5& $f_{0}+f_{1}$&26.65847&        29.39&  148.03&  26.64049&       28.76&  294.84&  26.65620&       29.27&  184.42&  26.65665&       29.72&  228.20\\
     6& $f_{1}-f_{0}$&3.40580&  25.47&  123.84&  3.42285&        27.36&  282.12& 3.40660&        24.36&  153.48&  3.40572&        25.44&  192.97\\
     7& $2f_{0}+f_{1}$&38.28217&        16.24&  84.60&   39.34215&       15.07&  150.52&  38.28280&       16.11&  101.50&  38.28216&       15.97&  123.94\\
     8& $4f_{0}$&46.51274&      12.80&  67.53&   46.55900&       13.02&  126.33&  46.50160&       12.91&  82.28&  46.50233&       12.99&  100.78\\
     9& $3f_{0}+f_{1}$&49.90705&        9.31&   49.30&   49.94062&       10.26&  97.83&   49.90820&       9.14&   57.64&   49.90796&       9.11&   71.07\\
    10& $5f_{0}$&58.13493&      6.23&   33.19&   57.15747&       4.19&  37.35&   58.12700&       5.89&   36.99&   58.12768&       5.99&   47.29\\
    11& $2f_{0}-f_{1}$&8.22789& 6.16&   29.75&   9.15509&        6.12&   66.30&  8.22257&        6.44&   40.27&   8.21896&        6.21&   46.75\\
    12& $4f_{0}+f_{1}$&61.54123&        4.99&   26.63&   60.33288&       5.85&   50.38&   61.53360&       5.15&   31.85&   61.53427&       4.92&   38.89\\
    13& $f_{0}+2f_{1}$&41.70315&        4.33&   22.78&   41.77526&       3.15&   31.22&   41.68940&       4.16&   26.42&   41.68853&       4.22&   32.67\\
    14& $2f_{1}$&30.08096&      3.77&   19.17&   30.06334&       5.03&   50.90&  30.06400&       3.32&   20.86&   30.06174&       4.00&   30.82\\
    15& $6f_{0}$&69.76912&      3.68&   19.56&   69.85913&       3.51&   28.13&   69.75240&       3.74&   23.61&   69.75322&       3.29&   26.59\\
    16& $2f_{1}-f_{0}$&19.81409&        3.45&   17.17&   18.22771&       3.46&   36.34&   19.84420&       3.04&   19.00&   19.84574&       3.16&   24.26\\
    17& $5f_{0}+f_{1}$&73.16342&        3.27&   17.40&   74.23048&       3.31&   25.21&   73.15900&       3.31&   20.83&   73.15980&       3.42&   27.88\\
    18& $6f_{0}+f_{1}$&84.76312&        2.15&   11.31&   84.77368&       2.83&   19.70&   84.78267&       2.21&   14.30&   84.78566&       2.19&   18.38\\
    19& $2f_{0}+2f_{1}$&53.34933&       2.07&   11.00&   53.69338&       1.82&   16.55&   53.31480&       1.66&   10.51&    53.31282&       2.26&   17.81\\
    20& $7f_{0}$&81.39280&      2.00&   10.53&    80.56729&       1.80&   12.95&   81.37984&       1.90&   12.21&   81.38032&       1.93&   15.96\\
    21& $7f_{0}+f_{1}$&96.39730&        1.45&   7.94&   96.36188&       1.51&   9.96&    96.40807&       1.60&   10.48&   96.41120&       1.62&   14.13\\
    22& $3f_{0}+2f_{1}$&64.93553&       1.60&   8.59&    63.83821&       3.22&   26.65&   64.94397&       1.64&   10.21&   64.93835&       1.59&   12.67\\
    23& $4f_{0}-f_{1}$&31.44828&        1.76&   9.02&    31.54795&       1.18&   11.96&   32.24212&       1.34&   8.43&    31.47003&       1.56&   12.05\\
    24& $2f_{1}-2f_{0}$&6.77661&        1.41&   6.81&   7.79420&        3.97&   42.67&  6.02561&        2.05&   12.85&    7.78861&        1.07&   8.06\\
    25& $2f_{1}-f_{0}$&18.45877&        1.40&   6.98&    18.22771&       3.46&   36.34&   18.44237&       2.04&   12.70&   18.43745&       1.93&   14.83\\
    26& $4f_{0}+2f_{1}$&76.54723&       1.32&   7.07&   ---&        ---&   ---&    75.79889&       1.16&   7.32&    76.56296&       1.18&   9.64\\
    27& $8f_{0}+f_{1}$&108.00750&       1.06&   5.94&   109.10478&      1.26&   7.68&    108.03347&      1.25&   8.15&    108.03549&      0.99&   8.77\\
    28& $8f_{0}$&93.03898&      1.22&   6.65&    ---&        ---&   ---&    93.01278&       0.88&   5.75&    93.00586&       1.18&   10.13\\
    29& $5f_{0}+2f_{1}$&88.20540&       1.05&   5.55&    ---&        ---&   ---&    88.18550&       1.06&   6.91&    88.18975&       1.05&   8.95\\
    30& $6f_{0}-f_{1}$&54.69265&        1.20&   6.36&    ---&        ---&   ---&    53.31480&       1.66&   10.51&    54.72485&       0.98&   7.74\\
    31& $3f_{1}$&44.05397&      1.03&   5.45&    ---&        ---&   ---&    43.09877&       1.13&   7.24&    43.09681&       1.00&   7.75\\
    32& $10f_{0}+f_{1}$&131.28786&      0.91&   5.43&    ---&        ---&   ---&    ---&        ---&   ---&    ---&        ---&   ---\\
    33& $9f_{0}+f_{1}$&119.62969&       0.82&   4.78&    118.71350&      0.97&   5.63&   119.65510&      0.72&   4.76&    118.65564&      0.57&   4.97\\
    34& $6f_{0}+2f_{1}$&100.82309&      0.78&   4.35&    100.77447&      0.99&   6.34&   99.81467&       1.01&   6.64&    99.81528&       0.77&   6.75\\
    35& $7f_{0}+2f_{1}$&110.50225&      0.73&   4.16&    110.99541&      0.78&   7.68&    111.43957&      0.69&   5.19&    ---&        ---&   ---\\
    36& $9f_{0}$&103.66567&     0.71&   4.01&    104.89839&      0.90&   5.74&    104.86428&      0.69&   4.48&    103.65346&      0.67&   5.94\\
    37& $8f_{0}+2f_{1}$&---&        ---&   ---&    ---&        ---&   ---& ---&        ---&   ---&    123.06760&      0.54&   4.82\\

\hline
\end{tabular}
}
\label{tab4}
\end{table*}

\begin{figure*}
\centering
\includegraphics[width=0.7\textwidth,height=0.4\textheight]{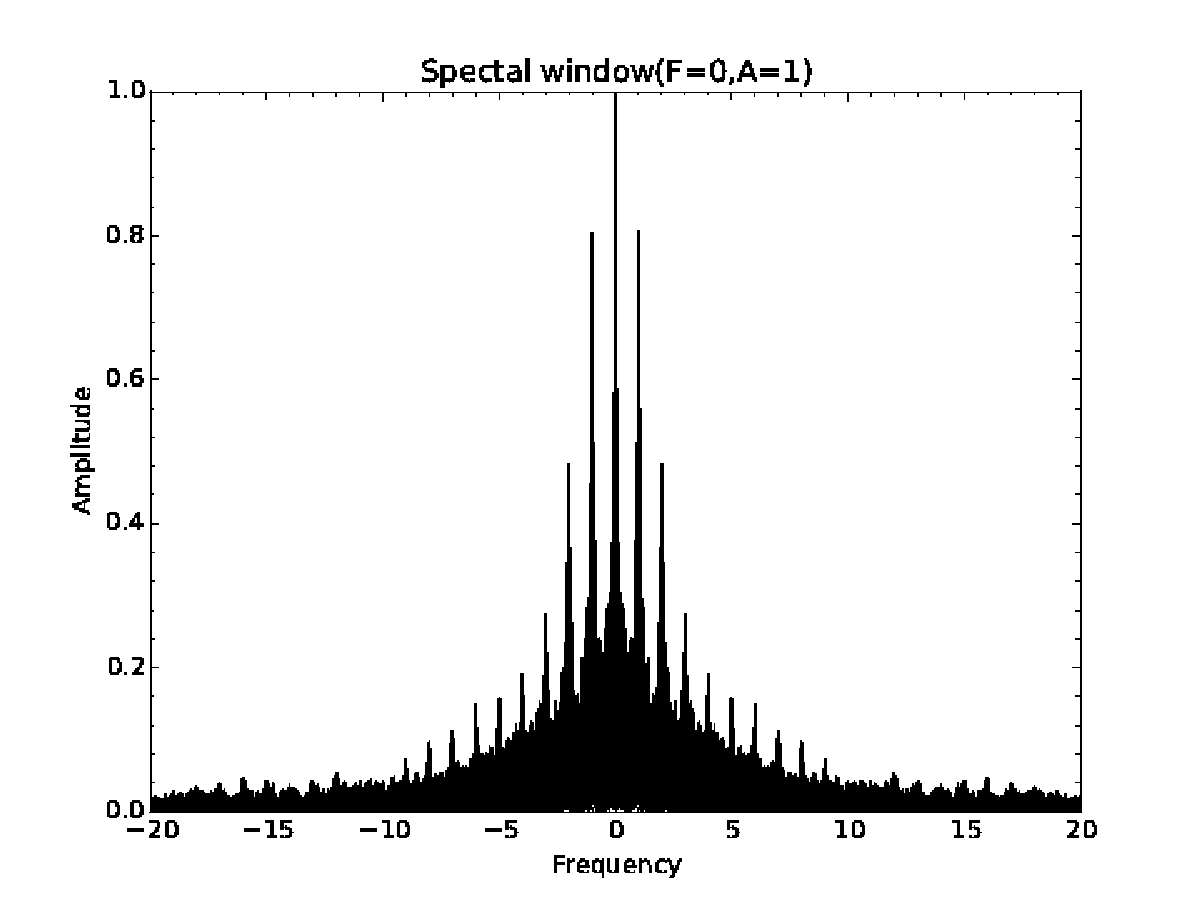}

\caption{Spectral window of the light curves in $V$ for AE UMa in 2009.}
\label{fig4}
\end{figure*}

\begin{figure*}
\centering
\includegraphics[width=1.0\textwidth,height=0.9\textheight]{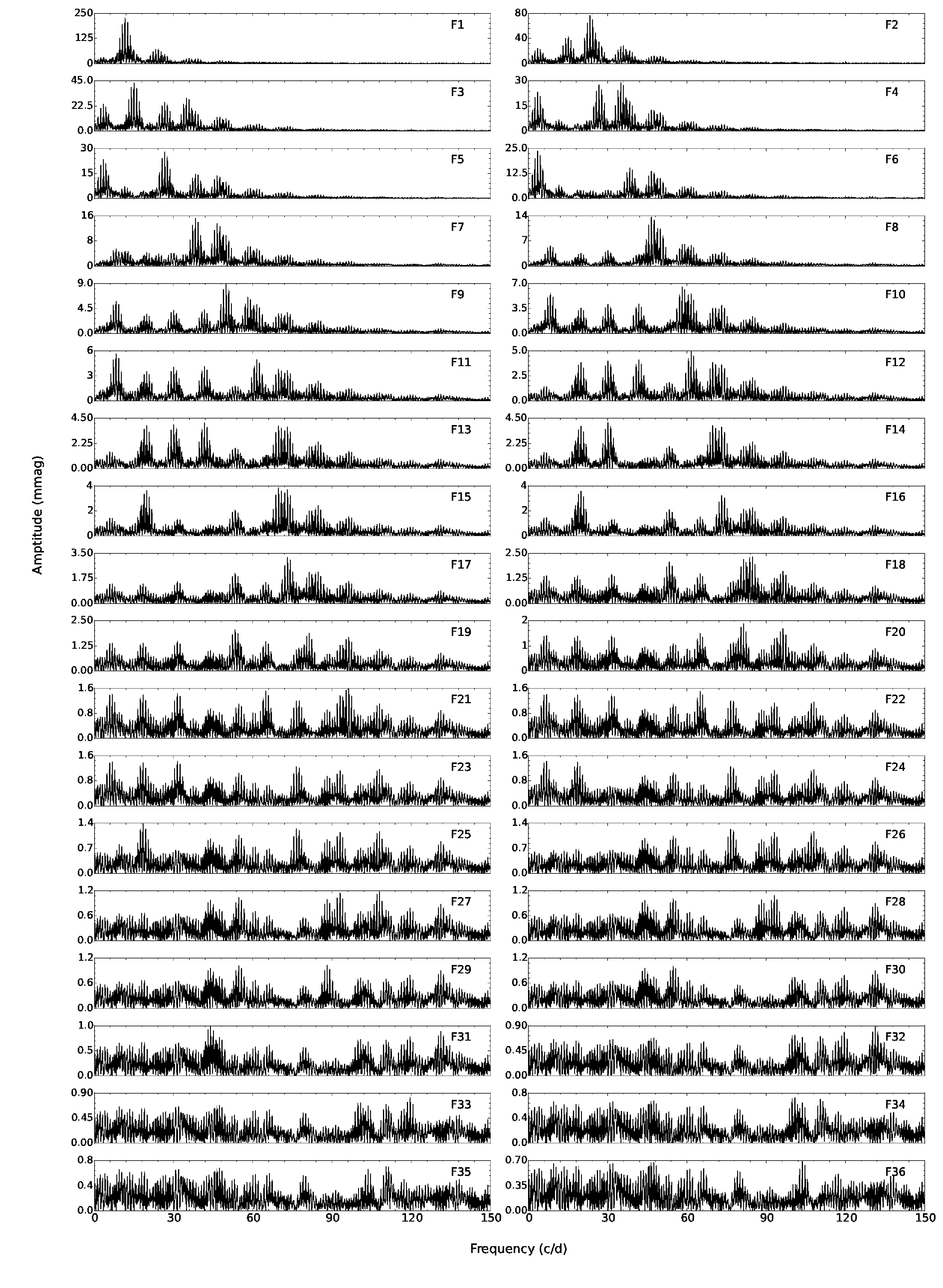}
\caption{Amplitude spectrum of the light curves in $V$ for AE UMa collected in 2009, and the amplitude spectra of the frequency pre-whitening process. Note that the y-axis scales are optimized concerning the highest peaks in the panels.}
\label{fig5}
\end{figure*}

Figure \ref{fig4} and Figure \ref{fig5} show the window function and the amplitude spectra of the frequency pre-whitening process for the light curves in $V$ in 2009, respectively.

As can be seen from Figure \ref{fig2}, the constructed curves fit well the light curves observed in 2009, 2010, 2011 and 2012, respectively, which shows that the fundamental and the first overtone frequencies, together with their harmonics and linear combinations, can explain the pulsation behavior of AE UMa.

To show the variations of the frequencies and amplitudes of pulsations of the star, we compare our results with those from \citet{Zhou2001}, which analysed the data for AE UMa from 1974 to 2001. By dividing the data into four segments of datasets following  \citet{Zhou2001}, we resolved the pulsation parameters of the fundamental and the first overtone frequencies of AE UMa and listed them in Table \ref{tab5}.

\begin{table*}
\caption{Frequencies and amplitudes of AE UMa for different segments of observations, including the data sets of \citet{Zhou2001} and our data. (The data of 1981-1987 has not been used due to its large scatters.)}
  \centering 

 \begin{tabular}{ccccc}
 \hline
  Years  & $f_{0}$  & $f_{1}$ & $A_{0}$  & $A_{1}$ \\

 \hline
1974-1977 & 11.62557 & 15.03097  & 216.9 & 34.1 \\
$^*$1981-1987 & 11.62290 & 15.07259  & 219.6 & 29.4 \\
1996-1998 & 11.62560 & 15.03122  & 210.9 & 36.8 \\
2000-2001 & 11.62561 & 15.03119  & 207.0 & 38.6 \\
2009-2012 & 11.62560 & 15.03123  & 219.0 & 45.1 \\
 \hline
Mean & 11.62560 & 15.03120 & 213.5 & 38.7 \\
$\sigma$ & 0.000015 & 0.000123 & 5.5 & 4.7 \\
\hline
 \end{tabular}
\label{tab5}
\end{table*}

\section{The $O-C$ Diagram}
With the new observations from 2009 to 2012, the light curves around the light maxima were fitted with a fourth polynomial by the nonlinear least-square method. The errors in polynomial fitting are consistent with the uncertainties estimated from Monte Carlo simulations of 100 iterations for each maximum. We obtained 84 times of maximum light in $V$ band as listed in Table \ref{tab6}.

\begin{table*}
\caption{Newly Determined Times of Maximum Light of AE UMa. $T_{max}$ is in $HJD - 2450000$. $\sigma$ is the estimated uncertainty of the times of maximum light in days.}
  \centering
    \scalebox{0.8}[0.6]
    { 
 \begin{tabular}{cccc}
   \hline
   \hline
 $T_{max}$  & $\sigma$  & $T_{max}$ & $\sigma$ \\
   \hline

4897.01876 & 0.00009 &  5621.11032 & 0.00017 \\
4897.10688 & 0.00008 &  5621.19860 & 0.00019 \\
4897.18828 & 0.00004 &  5937.39814 & 0.00012 \\
4898.04897 & 0.00010 &  5961.30507 & 0.00004 \\
4898.13312 & 0.00015 &  5961.39070 & 0.00010 \\
4898.99305 & 0.00013 &  5964.31479 & 0.00014 \\
4899.08369 & 0.00023 &  5964.40728 & 0.00019 \\
4899.17058 & 0.00013 &  5966.21331 & 0.00007 \\
4900.03182 & 0.00010 &  5966.29416 & 0.00005 \\
4900.11285 & 0.00008 &  5966.37965 & 0.00008 \\
4900.19779 & 0.00014 &  5967.33106 & 0.00009 \\
4901.05755 & 0.00027 &  5967.41592 & 0.00007 \\
4970.04314 & 0.00008 &  5968.27734 & 0.00007 \\
4972.10776 & 0.00024 &  5968.35817 & 0.00004 \\
5230.24553 & 0.00009 &  5969.30381 & 0.00007 \\
5230.33788 & 0.00010 &  5969.39594 & 0.00014 \\
5230.42007 & 0.00008 &  5997.00800 & 0.00008 \\
5231.28178 & 0.00008 &  5997.09064 & 0.00004 \\
5231.36352 & 0.00006 &  5997.17295 & 0.00004 \\
5583.43132 & 0.00012 &  5997.26391 & 0.00010 \\
5584.37869 & 0.00018 &  5998.03373 & 0.00008 \\
5585.41321 & 0.00008 &  5998.12209 & 0.00014 \\
5587.39457 & 0.00009 &  5998.21192 & 0.00008 \\
5588.42008 & 0.00008 &  5998.29274 & 0.00006 \\
5607.95282 & 0.00018 &  5998.98078 & 0.00009 \\
5608.03484 & 0.00011 &  5999.07267 & 0.00009 \\
5608.11800 & 0.00011 &  5999.15466 & 0.00005 \\
5608.20986 & 0.00022 &  5999.23746 & 0.00005 \\
5608.29535 & 0.00012 &  6000.01612 & 0.00005 \\
5608.37632 & 0.00011 &  6000.09780 & 0.00004 \\
5610.35912 & 0.00010 &  6000.18742 & 0.00009 \\
5611.38774 & 0.00016 &  6000.27618 & 0.00006 \\
5612.24692 & 0.00014 &  6001.04594 & 0.00006 \\
5612.33917 & 0.00022 &  6001.13682 & 0.00006 \\
5613.19799 & 0.00021 &  6001.21857 & 0.00003 \\
5616.98003 & 0.00011 &  6001.30228 & 0.00006 \\
5617.07053 & 0.00008 &  6062.03120 & 0.00005 \\
5617.15198 & 0.00005 &  6063.06910 & 0.00015 \\
5617.23583 & 0.00006 &  6064.01367 & 0.00008 \\
5617.32810 & 0.00010 &  6064.09479 & 0.00008 \\
5620.07717 & 0.00010 &  6065.04189 & 0.00009 \\
5620.16024 & 0.00017 &  6068.05799 & 0.00021 \\
   \hline
 \end{tabular}
 }
\label{tab6}
\end{table*}

In order to make $O-C$ analysis for the period change of AE UMa, we combined the new times of maximum light with those provided from previous literatures \footnote{\citet{199905IBVS}, \citet{Pocs2001}, \citet{2001IBVS}, \citet{Zhou2001}, \citet{2003IBVS}, \citet{200508IBVS}, \citet{200510IBVS}, \citet{200605IBVS}, \citet{200611IBVS}, \citet{200703IBVS}, \citet{2010JAVSO}, \citet{200710IBVS}, \citet{2009IBVS}, \citet{2010IBVS} and  \citet{2011BAVSM}}. We finally obtained 461 times of maximum light which are listed in Table \ref{tab7}.  We discarded 17 times of maximum light, that were collected with either photograph (pg) or visual (vis), with large uncertainties, comparing to those collected with CCD or photoelectric photometer (pe). We finally used 444 data points to {construct the $O-C$ (the Observed minus Calculated values) diagram}. The used linear ephemeris formula is,
\begin{equation}
  HJD_{\rm max} = 2442062.5824 + 0^{\rm d}.08601707 E
\end{equation}
 following \citet{Pocs2001}.\

\begin{figure*}
\centering
\includegraphics[width=0.8\hsize,height=0.5\hsize]{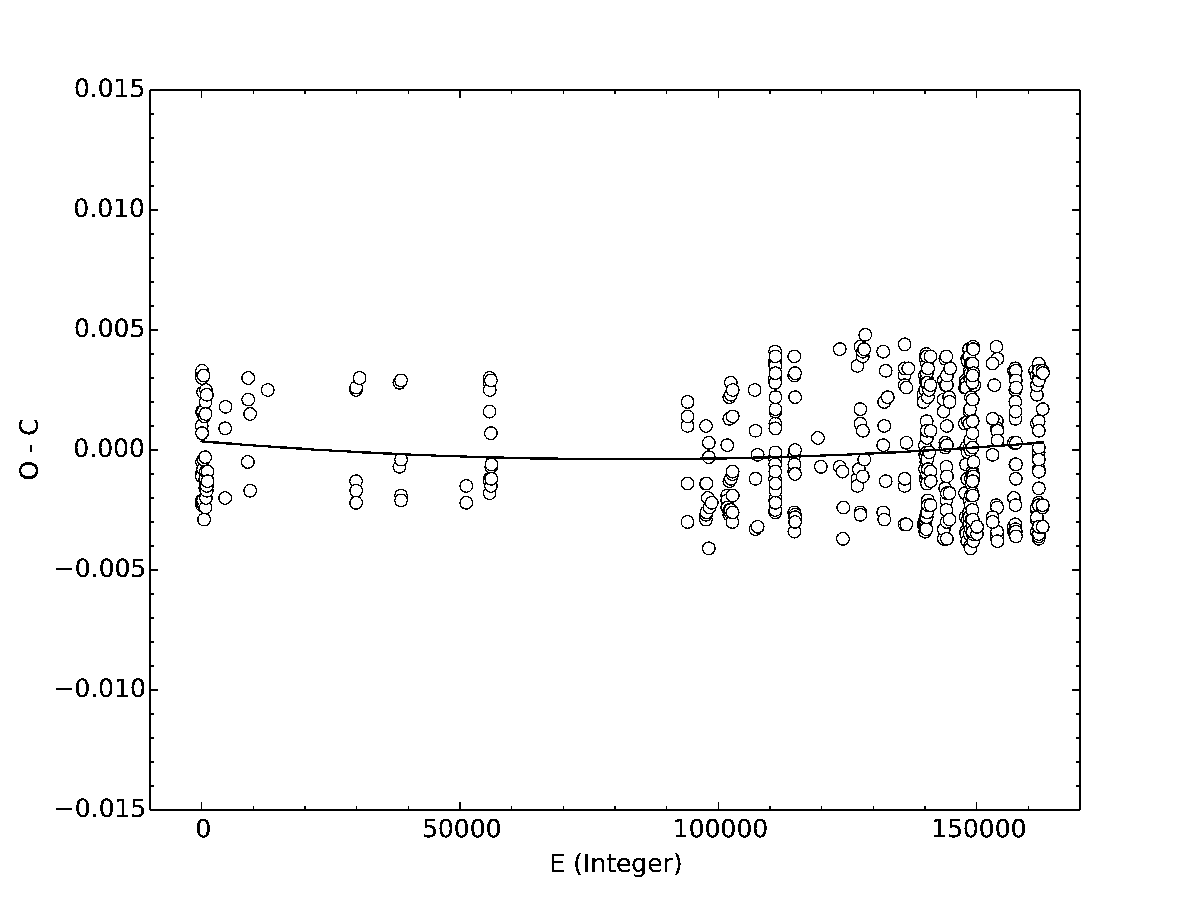}
\caption{$O-C$ diagram of AE UMa. The $O-C$ values are in $days$. E is the cycle number. The solid curve shows the fit concerning a continuous increasing period change.}
\label{fig6}
\end{figure*}

A linear fit to the 444 times of light maxima yields the ephemeris formula,
\begin{equation} \begin{split}
  HJD_{\rm max} &= 2442062.5818(\pm 0.0002)\\
              &+ 0^{\rm d}.086017078(\pm 0.000000002) E
\end{split} \end{equation}
 with a standard deviation of $\sigma_{0} = 0.00246$ days. The $O-C$ values are listed in Table \ref{tab7} as well. The $O-C$ diagram is shown in Figure \ref{fig6}.

\begin{deluxetable}{cccccc|cccccc}
\tabletypesize{\footnotesize}
\tablewidth{0pc}
\tablecaption{Times of Maximum Light and $O-C$ values of AE UMa. $T_{max}$ is the observed times of maximum light in $HJD-2400000$. $E$: Cycle number. $O-C$ is in days. Det: detector (pg=photograph, vis=visual, pe=photoelectric photometer). Points not used in the $O-C$ analysis are marked with asterisk.}
\tablehead{
\colhead{NO.} & \colhead{$T_{max}$} & \colhead{\emph{E}} & \colhead{$O-C$} & \colhead{Det} & \colhead{S} &
\colhead{NO.} & \colhead{$T_{max}$} & \colhead{\emph{E}} & \colhead{$O-C$} & \colhead{Det} & \colhead{S} }

\startdata
1&  28632.398&  -156133&  ---&  pg&  (1)*&      26& 42087.5263& 290& -0.000808& pe& (4)\\
2&  31875.122&  -118434&  ---&  pg&  (2)*&      27& 42087.6155& 291& 0.002375& pe& (4)\\
3&  33379.256&  -100948&  ---&  pg&  (2)*&      28& 42095.5298& 383& 0.003105& pe& (3)\\
4&  35601.188&  -75117&  ---&  pg&  (2)*&       29& 42095.6123& 384& -0.000412& pe& (3)\\
5&  35604.337&  -75080&  ---&  vis&  (1)*&      30& 42103.3513& 474& -0.002947& pe& (4)\\
6&  35607.173&  -75047&  ---&  pg&  (2)*&       31& 42106.4523& 510& 0.001439& pe& (3)\\
7&  35981.202&  -70699&  ---&  pg&  (2)*&       32& 42119.5252& 662& -0.000255& pe& (3)\\
8&  38106.402&  -45992&  ---&  vis&  (1)*&      33& 42121.5025& 685& -0.001347& pe& (3)\\
9&  41059.368&  -11662&  ---&  vis&  (1)*&      34& 42122.3628& 695& -0.001218& pe& (4)\\
10&  41773.223&  -3363&  ---&  vis&  (1)*&      35& 42122.4484& 696& -0.001635& pe& (4)\\
11& 42062.5832& 0& 0.001039& pe& (3)&   36& 42128.2968& 764& -0.002395& pe& (3)\\
12& 42065.5959& 35& 0.003142& pe& (4)&  37& 42128.3872& 765& 0.001988& pe& (3)\\
13& 42065.6778& 36& -0.000975& pe& (4)& 38& 42128.4727& 766& 0.001471& pe& (3)\\
14& 42068.3432& 67& -0.002104& pe& (4)& 39& 42128.5557& 767& -0.001546& pe& (3)\\
15& 42068.4302& 68& -0.001121& pe& (4)& 40& 42133.4627& 824& 0.002482& pe& (3)\\
16& 42068.5203& 69& 0.002962& pe& (4)&  41& 42133.5442& 825& -0.002035& pe& (3)\\
17& 42068.6029& 70& -0.000455& pe& (4)& 42& 42134.4055& 835& -0.000906& pe& (3)\\
18& 42068.6871& 71& -0.002272& pe& (4)& 43& 42147.3933& 986& -0.001682& pe& (3)\\
19& 42069.3808& 79& 0.003292& pe& (4)&  44& 42148.4295& 998& 0.002313& pe& (3)\\
20& 42069.4651& 80& 0.001574& pe& (4)&  45& 42148.5117& 999& -0.001504& pe& (3)\\
21& 42069.5473& 81& -0.002243& pe& (4)& 46& 42159.4365& 1126& -0.000871& pe& (3)\\
22& 42069.6363& 82& 0.000740& pe& (4)&  47& 42161.4145& 1149& -0.001263& pe& (3)\\
23& 42086.4965& 278& 0.001597& pe& (4)& 48& 42453.5306& 4545& 0.000899& pe& (3)\\
24& 42086.5787& 279& -0.002221& pe& (4)&        49& 42453.6137& 4546& -0.002018& pe& (3)\\
25& 42087.4390& 289& -0.002091& pe& (4)&        50& 42460.4989& 4626& 0.001817& pe& (3)\\
51&  42532.407&  5462&  ---&  vis&  (5)*&       76& 46468.4601& 51221& -0.002180& pe& (3)\\
52& 42830.6280& 8929& -0.000499& pe& (3)&       77& 46468.5468& 51222& -0.001497& pe& (3)\\
53& 42837.5120& 9009& 0.002136& pe& (3)&        78& 46855.6279& 55722& 0.002780& pe& (8)\\
54& 42838.4591& 9020& 0.003049& pe& (3)&        79& 46856.5729& 55733& 0.001592& pe& (8)\\
55&  42866.496&  9346&  ---&  vis&  (5)*&       80& 46856.6561& 55734& -0.001225& pe& (8)\\
56& 42869.3377& 9379& 0.001523& pe& (3)&        81& 46857.6017& 55745& -0.001812& pe& (8)\\
57& 42869.4205& 9380& -0.001694& pe& (3)&       82& 46857.6925& 55746& 0.002970& pe& (8)\\
58& 43162.5708& 12788& 0.002457& pe& (3)&       83& 46858.6382& 55757& 0.002483& pe& (8)\\
59& 44633.4626& 29888& 0.002451& pe& (3)&       84& 46859.6666& 55769& -0.001322& pe& (8)\\
60& 44633.5440& 29889& -0.002166& pe& (3)&      85& 46878.4181& 55987& -0.001544& pe& (8)\\
61& 44633.6309& 29890& -0.001283& pe& (3)&      86& 46878.5064& 55988& 0.000739& pe& (8)\\
62& 44634.4046& 29899& -0.001737& pe& (3)&      87& 46878.5946& 55989& 0.002922& pe& (8)\\
63& 44634.4902& 29900& -0.002154& pe& (3)&      88& 46884.5262& 56058& -0.000656& pe& (8)\\
64& 44634.5810& 29901& 0.002629& pe& (3)&       89& 46884.6117& 56059& -0.001173& pe& (8)\\
65& 44692.4709& 30574& 0.003043& pe& (3)&       90& 46886.5907& 56082& -0.000566& pe& (8)\\
66&  44696.343&  30619&  ---&  vis&  (6)*&      91&  48683.317&  76970&  ---&  vis&  (9)*\\
67&  44696.426&  30620&  ---&  vis&  (6)*&      92& 50151.4564& 94038& 0.000981& pe& (3)\\
68&  44696.520&  30621&  ---&  vis&  (6)*&      93& 50151.5384& 94039& -0.003036& pe& (3)\\
69& 45355.4902& 38282& 0.002786& pe& (3)&       94& 50152.3170& 94048& 0.001411& pe& (3)\\
70& 45355.5727& 38283& -0.000731& pe& (3)&      95& 50152.4862& 94050& -0.001424& pe& (3)\\
71& 45382.3228& 38594& -0.001939& pe& (3)&      96& 50152.5756& 94051& 0.001959& pe& (3)\\
72& 45382.4104& 38595& -0.000356& pe& (3)&      97& 50458.8815& 97612& 0.001034& CCD& (10)\\
73& 45382.4997& 38596& 0.002927& pe& (3)&       98& 50458.9636& 97613& -0.002883& CCD& (10)\\
74& 45382.5807& 38597& -0.002090& pe& (3)&      99& 50459.8240& 97623& -0.002654& CCD& (10)\\
75&  46114.332&  47104&  ---&  vis&  (7)*&      100& 50459.9113& 97624& -0.001371& CCD& (10)\\
101& 50467.7388& 97715& -0.001425& CCD& (10)&   126& 50902.2976& 102768& -0.000923& pe& (3)\\
102& 50467.8236& 97716& -0.002642& CCD& (10)&   127& 50902.3819& 102779& -0.002640& pe& (3)\\
103& 50490.3607& 97978& -0.002018& pe& (3)&     128& 50903.3321& 102780& 0.001372& pe& (3)\\
104& 50505.6697& 98156& -0.004058& CCD& (10)&   129& 50903.4192& 102781& 0.002455& pe& (3)\\
105& 50505.7595& 98157& -0.000275& CCD& (10)&   130& 50903.5009& 107036& -0.001862& CCD& (13)\\
106& 50505.8461& 98158& 0.000308& CCD& (10)&    131& 51269.5080& 107198& 0.002550& CCD& (13)\\
107& 50516.7676& 98285& -0.002362& CCD& (10)&   132& 51283.4410& 107199& 0.000782& CCD& (13)\\
108& 50554.4432& 98723& -0.002243& pe& (3)&     133& 51283.5250& 107200& -0.001235& CCD& (13)\\
109& 50813.3550& 101733& -0.001860& pe& (3)&    134& 51283.6090& 107604& -0.003252& CCD& (13)\\
110& 50813.4408& 101734& -0.002077& pe& (3)&    135& 51318.3630& 107605& -0.000154& CCD& (13)\\
111& 50813.6151& 101736& 0.000189& pe& (3)&     136& 51318.4460& 110972& -0.003171& CCD& (14)\\
112& 50813.6985& 101737& -0.002428& pe& (3)&    137& 51608.0716& 110973& 0.002908& CCD& (14)\\
113& 50848.4540& 102141& 0.002170& pe& (3)&     138& 51608.1577& 110974& 0.002991& CCD& (14)\\
114& 50848.5391& 102142& 0.001253& pe& (3)&     139& 51608.2395& 110975& -0.001226& CCD& (14)\\
115& 50848.6212& 102143& -0.002664& pe& (3)&    140& 51608.3264& 110983& -0.000343& CCD& (14)\\
116& 50849.4815& 102153& -0.002535& pe& (3)&    141& 51609.0186& 110984& 0.003720& CCD& (14)\\
117& 50849.5688& 102154& -0.001252& pe& (3)&    142& 51609.1006& 110985& -0.000297& CCD& (14)\\
118& 50862.3840& 102303& -0.002597& CCD& (13)&  143& 51609.1865& 110986& -0.000414& CCD& (14)\\
119& 50862.3840& 102418& -0.002597& pe& (3)&    144& 51609.2770& 110987& 0.004069& CCD& (14)\\
120& 50872.2809& 102419& 0.002339& pe& (3)&     145& 51609.3583& 110995& -0.000648& CCD& (14)\\
121& 50872.3634& 102420& -0.001178& pe& (3)&    146& 51610.0450& 111006& -0.002085& CCD& (14)\\
122& 50872.4481& 102421& -0.002496& pe& (3)&    147& 51610.9969& 111007& 0.003627& CCD& (14)\\
123& 50872.5394& 102733& 0.002787& pe& (3)&     148& 51611.0821& 111008& 0.002810& CCD& (14)\\
124& 50899.3729& 102734& -0.001042& pe& (3)&    149& 51611.1627& 111018& -0.002607& CCD& (14)\\
125& 50899.4570& 102767& -0.002959& pe& (3)&    150& 51612.0246& 111019& -0.000878& CCD& (14)\\
151& 51612.1090& 111020& -0.002495& CCD& (14)&  176& 51941.2979& 114847& -0.000978& CCD& (14)\\
152& 51612.2010& 111021& 0.003488& CCD& (14)&   177& 51941.3881& 114856& 0.003205& CCD& (14)\\
153& 51612.2846& 111022& 0.001071& CCD& (14)&   178& 51942.1562& 114857& -0.002849& CCD& (14)\\
154& 51612.3704& 111029& 0.000854& CCD& (14)&   179& 51942.2473& 114858& 0.002234& CCD& (14)\\
155& 51612.9692& 111030& -0.002466& CCD& (14)&  180& 51942.3311& 114859& 0.000017& CCD& (14)\\
156& 51613.0609& 111031& 0.003217& CCD& (14)&   181& 51942.4141& 119265& -0.003000& CCD& (15)\\
157& 51613.1453& 111032& 0.001600& CCD& (14)&   182& 52321.4089& 119846& 0.000521& CCD& (15)\\
158& 51613.2276& 111033& -0.002117& CCD& (14)&  183& 52371.3836& 123496& -0.000706& CCD& (15)\\
159& 51613.3156& 111053& -0.000134& CCD& (14)&  184& 52685.3460& 123497& -0.000673& CCD& (15)\\
160& 51615.0341& 111054& -0.001976& CCD& (14)&  185& 52685.4369& 124020& 0.004210& CCD& (15)\\
161& 51615.1260& 111055& 0.003907& CCD& (14)&   186& 52730.4187& 124124& -0.000926& CCD& (15)\\
162& 51615.2098& 111056& 0.001690& CCD& (14)&   187& 52739.3617& 124195& -0.003703& CCD& (16)\\
163& 51615.2919& 111064& -0.002227& CCD& (14)&  188& 52745.4702& 126928& -0.002416& CCD& (17)\\
164& 51615.9855& 111065& 0.003236& CCD& (14)&   189& 52980.5608& 126929& 0.003484& CCD& (17)\\
165& 51616.0705& 111066& 0.002219& CCD& (14)&   190& 52980.6421& 126930& -0.001233& CCD& (17)\\
166& 51616.1526& 111067& -0.001698& CCD& (14)&  191& 52980.7279& 127195& -0.001451& CCD& (16)\\
167& 51616.2389& 114706& -0.001415& CCD& (14)&  192& 53003.5231& 127483& -0.000779& CCD& (16)\\
168& 51929.2556& 114707& -0.000886& CCD& (14)&  193& 53028.2942& 127484& -0.002600& CCD& (16)\\
169& 51929.3464& 114717& 0.003897& CCD& (14)&   194& 53028.3871& 127485& 0.004283& CCD& (16)\\
170& 51930.2058& 114718& 0.003126& CCD& (14)&   195& 53028.4705& 127486& 0.001666& CCD& (16)\\
171& 51930.2885& 114719& -0.000191& CCD& (14)&  196& 53028.5522& 127487& -0.002651& CCD& (16)\\
172& 51930.3721& 114729& -0.002608& CCD& (14)&  197& 53028.6420& 127961& 0.001132& CCD& (16)\\
173& 51931.2315& 114730& -0.003379& CCD& (14)&  198& 53069.4119& 127962& -0.001068& CCD& (16)\\
174& 51931.3203& 114845& -0.000596& CCD& (14)&  199& 53069.5029& 127973& 0.003915& CCD& (16)\\
175& 51941.2102& 114846& -0.002661& CCD& (14)&  200& 53070.4493& 127974& 0.004127& CCD& (16)\\
201& 53070.5320& 128205& 0.000810& CCD& (16)&   226& 54079.8580& 139709& 0.002285& CCD& (21)\\
202& 53090.4053& 128251& 0.004162& CCD& (16)&   227& 54079.9437& 139973& 0.001967& CCD& (21)\\
203& 53094.3575& 128437& -0.000424& CCD& (19)&  228& 54102.6471& 139974& -0.003144& CCD& (21)\\
204& 53110.3619& 131915& 0.004798& CCD& (18)&   229& 54102.7365& 139975& 0.000238& CCD& (21)\\
205& 53409.5286& 131916& 0.004064& CCD& (18)&   230& 54102.8254& 139985& 0.003121& CCD& (21)\\
206& 53409.6108& 131917& 0.000247& CCD& (18)&   231& 54103.6849& 139986& 0.002450& CCD& (21)\\
207& 53409.6940& 132122& -0.002570& CCD& (19)&  232& 54103.7677& 139997& -0.000767& CCD& (21)\\
208& 53427.3272& 132123& -0.002873& CCD& (19)&  233& 54104.7117& 140031& -0.002955& CCD& (21)\\
209& 53427.4181& 132124& 0.002010& CCD& (19)&   234& 54107.6358& 140032& -0.003436& CCD& (21)\\
210& 53427.5031& 132402& 0.000993& CCD& (17)&   235& 54107.7251& 140033& -0.000153& CCD& (21)\\
211& 53451.4136& 132403& -0.001258& CCD& (17)&  236& 54107.8138& 140034& 0.002530& CCD& (21)\\
212& 53451.5042& 132785& 0.003325& CCD& (17)&   237& 54107.8944& 140067& -0.002887& CCD& (21)\\
213& 53484.3616& 136052& 0.002197& CCD& (20)&   238& 54110.7388& 140068& 0.002949& CCD& (21)\\
214& 53765.3803& 136053& 0.003066& CCD& (20)&   239& 54110.8207& 140069& -0.001168& CCD& (21)\\
215& 53765.4660& 136054& 0.002749& CCD& (20)&   240& 54110.9051& 140070& -0.002785& CCD& (21)\\
216& 53765.5462& 136063& -0.003068& CCD& (20)&  241& 54110.9977& 140241& 0.003798& CCD& (21)\\
217& 53766.3278& 136064& 0.004378& CCD& (20)&   242& 54125.7056& 140242& 0.002775& CCD& (21)\\
218& 53766.4079& 136065& -0.001539& CCD& (20)&  243& 54125.7883& 140243& -0.000542& CCD& (21)\\
219& 53766.4943& 136066& -0.001156& CCD& (20)&  244& 54125.8716& 140244& -0.003259& CCD& (21)\\
220& 53766.5849& 136389& 0.003427& CCD& (18)&   245& 54125.9649& 140309& 0.004024& CCD& (21)\\
221& 53794.3619& 136400& -0.003093& CCD& (18)&  246& 54131.5550& 140310& 0.003013& CCD& (21)\\
222& 53795.3115& 136401& 0.000319& CCD& (18)&   247& 54131.6392& 140311& 0.001196& CCD& (21)\\
223& 53795.3998& 136776& 0.002602& CCD& (18)&   248& 54131.7212& 140312& -0.002821& CCD& (21)\\
224& 53827.6570& 139707& 0.003393& CCD& (21)&   249& 54131.8106& 140313& 0.000562& CCD& (21)\\
225& 54079.7666& 139708& -0.003098& CCD& (21)&  250& 54131.8989& 140367& 0.002845& CCD& (21)\\
251& 54136.5446& 140368& 0.003622& CCD& (21)&   276& 54198.3850& 141086& -0.002266& CCD& (22)\\
252& 54136.6275& 140369& 0.000505& CCD& (21)&   277& 54198.4741& 141087& 0.000817& CCD& (22)\\
253& 54136.7104& 140370& -0.002612& CCD& (21)&  278& 54202.4288& 141133& -0.001270& CCD& (22)\\
254& 54136.7998& 140371& 0.000771& CCD& (21)&   279& 54414.8939& 143603& 0.001615& CCD& (21)\\
255& 54136.8882& 140372& 0.003154& CCD& (21)&   280& 54414.9804& 143604& 0.002098& CCD& (21)\\
256& 54136.9697& 140380& -0.001364& CCD& (21)&  281& 54417.7271& 143636& -0.003749& CCD& (21)\\
257& 54137.6581& 140381& -0.001100& CCD& (21)&  282& 54417.9058& 143638& 0.002917& CCD& (21)\\
258& 54137.7491& 140382& 0.003883& CCD& (21)&   283& 54417.9856& 143639& -0.003300& CCD& (21)\\
259& 54137.8311& 140404& -0.000134& CCD& (21)&  284& 54440.6958& 143903& -0.001612& CCD& (21)\\
260& 54139.7226& 140405& -0.001010& CCD& (21)&  285& 54442.7619& 143927& 0.000078& CCD& (21)\\
261& 54139.8124& 140406& 0.002772& CCD& (21)&   286& 54442.8516& 143928& 0.003761& CCD& (21)\\
262& 54139.8942& 140428& -0.001445& CCD& (21)&  287& 54442.9342& 143929& 0.000344& CCD& (21)\\
263& 54141.7876& 140429& -0.000421& CCD& (21)&  288& 54451.6243& 144030& 0.002717& CCD& (21)\\
264& 54141.8768& 140430& 0.002762& CCD& (21)&   289& 54460.6513& 144135& -0.002077& CCD& (21)\\
265& 54141.9575& 140554& -0.002555& CCD& (21)&  290& 54460.7433& 144136& 0.003906& CCD& (21)\\
266& 54152.6241& 140565& -0.002074& CCD& (21)&  291& 54460.8247& 144137& -0.000712& CCD& (21)\\
267& 54153.5746& 140578& 0.002238& CCD& (21)&   292& 54460.9089& 144138& -0.002529& CCD& (21)\\
268& 54154.6883& 140579& -0.002284& CCD& (21)&  293& 54467.7955& 144218& 0.002704& CCD& (21)\\
269& 54154.7800& 140580& 0.003399& CCD& (21)&   294& 54468.7392& 144229& 0.000216& CCD& (21)\\
270& 54154.8618& 140773& -0.000818& CCD& (22)&  295& 54468.8220& 144230& -0.003001& CCD& (21)\\
271& 54171.4664& 140808& 0.002483& CCD& (22)&   296& 54468.9120& 144231& 0.000982& CCD& (21)\\
272& 54175.4206& 140819& -0.000103& CCD& (22)&  297& 54469.6815& 144240& -0.003672& CCD& (21)\\
273& 54196.4976& 141064& 0.002710& CCD& (22)&   298& 54469.7701& 144241& -0.001089& CCD& (21)\\
274& 54197.3590& 141074& 0.003939& CCD& (22)&   299& 54469.8603& 144242& 0.003094& CCD& (21)\\
275& 54197.4402& 141075& -0.000878& CCD& (22)&  300& 54469.9414& 144243& -0.001823& CCD& (21)\\
301& 54506.5887& 144669& 0.002196& CCD& (23)&   326& 54837.7565& 148519& 0.004191& CCD& (21)\\
302& 54512.5199& 144738& -0.001784& CCD& (23)&  327& 54843.6886& 148588& 0.001112& CCD& (21)\\
303& 54513.4650& 144749& -0.002872& CCD& (23)&  328& 54843.7704& 148589& -0.003105& CCD& (21)\\
304& 54513.5559& 144750& 0.002011& CCD& (23)&   329& 54843.8567& 148590& -0.002822& CCD& (21)\\
305& 54524.4815& 144877& 0.003441& CCD& (23)&   330& 54843.9494& 148591& 0.003861& CCD& (21)\\
306& 54769.8859& 147730& 0.001077& CCD& (21)&   331& 54846.6137& 148622& 0.001631& CCD& (21)\\
307& 54770.8336& 147741& 0.002589& CCD& (21)&   332& 54846.6960& 148623& -0.002086& CCD& (21)\\
308& 54770.9152& 147742& -0.001828& CCD& (21)&  333& 54847.7336& 148635& 0.003309& CCD& (21)\\
309& 54781.8384& 147869& -0.002798& CCD& (21)&  334& 54847.8164& 148636& 0.000092& CCD& (21)\\
310& 54781.9301& 147870& 0.002884& CCD& (21)&   335& 54847.8988& 148637& -0.003525& CCD& (21)\\
311& 54788.8080& 147950& -0.000583& CCD& (21)&  336& 54855.6469& 148727& 0.003036& CCD& (21)\\
312& 54788.8912& 147951& -0.003400& CCD& (21)&  337& 54855.7316& 148728& 0.001719& CCD& (21)\\
313& 54788.9833& 147952& 0.002683& CCD& (21)&   338& 54855.8119& 148729& -0.003998& CCD& (21)\\
314& 54791.7333& 147984& 0.000136& CCD& (21)&   339& 54855.9023& 148730& 0.000385& CCD& (21)\\
315& 54791.8157& 147985& -0.003481& CCD& (21)&  340& 54864.5873& 148831& -0.002341& CCD& (21)\\
316& 54791.9080& 147986& 0.002802& CCD& (21)&   341& 54864.6724& 148832& -0.003258& CCD& (21)\\
317& 54791.9938& 147987& 0.002585& CCD& (21)&   342& 54864.7649& 148833& 0.003224& CCD& (21)\\
318& 54807.8222& 148171& 0.003840& CCD& (21)&   343& 54864.8477& 148834& 0.000007& CCD& (21)\\
319& 54807.9032& 148172& -0.001177& CCD& (21)&  344& 54864.9296& 148835& -0.004110& CCD& (21)\\
320& 54807.9866& 148173& -0.003795& CCD& (21)&  345& 54868.6346& 148878& 0.002155& CCD& (21)\\
321& 54816.6793& 148274& 0.001179& CCD& (21)&   346& 54868.7166& 148879& -0.001862& CCD& (21)\\
322& 54816.7612& 148275& -0.002938& CCD& (21)&  347& 54868.8033& 148880& -0.001179& CCD& (21)\\
323& 54816.8501& 148276& -0.000055& CCD& (21)&  348& 54868.8941& 148881& 0.003604& CCD& (21)\\
324& 54816.9399& 148277& 0.003728& CCD& (21)&   349& 54878.6935& 148995& -0.002945& CCD& (21)\\
325& 54837.6637& 148518& -0.002591& CCD& (21)&  350& 54878.7811& 148996& -0.001362& CCD& (21)\\
351& 54878.8721& 148997& 0.003621& CCD& (21)&   376& 54924.3742& 149526& 0.003777& CCD& (24)\\
352& 54894.4412& 149178& 0.003628& CCD& (24)&   377& 54970.0431& 150057& 0.000360& CCD& (26)\\
353& 54894.5227& 149179& -0.000890& CCD& (24)&  378& 54972.1078& 150081& 0.000715& CCD& (26)\\
354& 54894.6071& 149180& -0.002507& CCD& (24)&  379& 55230.2455& 153082& 0.002798& CCD& (26)\\
355& 54897.0188& 149208& 0.000058& CCD& (26)&   380& 55230.3379& 153083& -0.001819& CCD& (26)\\
356& 54897.1069& 149209& 0.004345& CCD& (26)&   381& 55230.4201& 153084& -0.001290& CCD& (26)\\
357& 54897.1883& 149210& -0.000972& CCD& (26)&  382& 55231.2818& 153094& -0.003207& CCD& (26)\\
358& 54898.0490& 149220& -0.003829& CCD& (26)&  383& 55231.3635& 153095& -0.003478& CCD& (26)\\
359& 54898.1331& 149221& 0.004153& CCD& (26)&   384& 55259.4108& 153421& 0.001205& CCD& (25)\\
360& 54898.3084& 149223& -0.001064& CCD& (24)&  385& 55293.3826& 153816& 0.002087& CCD& (25)\\
361& 54898.9930& 149231& -0.003252& CCD& (26)&  386& 55293.4752& 153817& 0.003117& CCD& (25)\\
362& 54899.0837& 149232& 0.003172& CCD& (26)&   387& 55302.3318& 153920& -0.001901& CCD& (25)\\
363& 54899.1706& 149233& -0.000545& CCD& (26)&  388& 55303.3591& 153932& -0.002918& CCD& (25)\\
364& 54900.0318& 149243& 0.002679& CCD& (26)&   389& 55304.3959& 153944& -0.003389& CCD& (25)\\
365& 54900.1128& 149244& 0.002702& CCD& (26)&   390& 55304.4775& 153945& -0.003497& CCD& (25)\\
366& 54900.1978& 149245& -0.002250& CCD& (26)&  391& 55305.3388& 153955& -0.003207& CCD& (25)\\
367& 54901.0575& 149255& 0.004333& CCD& (26)&   392& 55305.4238& 153956& -0.002804& CCD& (25)\\
368& 54904.4199& 149294& 0.001172& CCD& (24)&   393& 55309.3848& 154002& 0.003579& CCD& (25)\\
369& 54904.5006& 149295& -0.003733& CCD& (24)&  394& 55310.4124& 154014& -0.000238& CCD& (25)\\
370& 54909.3147& 149351& 0.000862& CCD& (24)&   395& 55311.3662& 154025& 0.001291& CCD& (25)\\
371& 54909.4087& 149352& -0.003555& CCD& (24)&  396& 55311.4488& 154026& -0.003026& CCD& (25)\\
372& 54909.4895& 149353& -0.002426& CCD& (24)&  397& 55583.4313& 157188& -0.003191& CCD& (26)\\
373& 54910.4335& 149364& -0.003443& CCD& (24)&  398& 55584.3787& 157199& -0.001979& CCD& (26)\\
374& 54912.3323& 149386& 0.000770& CCD& (24)&   399& 55585.4132& 157211& 0.000316& CCD& (26)\\
375& 54912.4146& 149387& -0.003835& CCD& (24)&  400& 55587.3946& 157234& 0.003323& CCD& (26)\\
401& 55588.4201& 157246& -0.003383& CCD& (26)&  426& 55964.4073& 161617& 0.003096& CCD& (26)\\
402& 55607.9528& 157473& 0.003437& CCD& (26)&   427& 55966.2133& 161638& 0.002737& CCD& (26)\\
403& 55608.0348& 157474& -0.000580& CCD& (26)&  428& 55966.2942& 161639& -0.002380& CCD& (26)\\
404& 55608.1180& 157475& -0.003397& CCD& (26)&  429& 55966.3797& 161640& -0.002897& CCD& (26)\\
405& 55608.2099& 157476& 0.002486& CCD& (26)&   430& 55967.3311& 161651& 0.002315& CCD& (26)\\
406& 55608.2954& 157477& 0.001969& CCD& (26)&   431& 55967.4159& 161652& 0.001098& CCD& (26)\\
407& 55608.3763& 157478& -0.003148& CCD& (26)&  432& 55968.2773& 161662& 0.002327& CCD& (26)\\
408& 55610.3591& 157501& 0.001258& CCD& (26)&   433& 55968.3582& 161663& -0.002790& CCD& (26)\\
409& 55611.3877& 157513& -0.002347& CCD& (26)&  434& 55969.3038& 161674& -0.003379& CCD& (26)\\
410& 55612.2469& 157523& -0.003318& CCD& (26)&  435& 55969.3959& 161675& 0.002704& CCD& (26)\\
411& 55612.3392& 157524& 0.002965& CCD& (26)&   436& 55997.0080& 161996& 0.003317& CCD& (26)\\
412& 55613.1980& 157534& 0.001594& CCD& (26)&   437& 55997.0906& 161997& -0.000100& CCD& (26)\\
413& 55616.9800& 157578& -0.001158& CCD& (26)&  438& 55997.1730& 161998& -0.003717& CCD& (26)\\
414& 55617.0705& 157579& 0.003325& CCD& (26)&   439& 55997.2639& 161999& 0.001166& CCD& (26)\\
415& 55617.1520& 157580& -0.001192& CCD& (26)&  440& 55998.0337& 162008& -0.003188& CCD& (26)\\
416& 55617.2358& 157581& -0.003409& CCD& (26)&  441& 55998.1221& 162009& -0.000805& CCD& (26)\\
417& 55617.3281& 157582& 0.002874& CCD& (26)&   442& 55998.2119& 162010& 0.002977& CCD& (26)\\
418& 55620.0772& 157614& -0.000573& CCD& (26)&  443& 55998.2927& 162011& -0.002240& CCD& (26)\\
419& 55620.1602& 157615& -0.003590& CCD& (26)&  444& 55998.9808& 162019& -0.002276& CCD& (26)\\
420& 55621.1103& 157626& 0.000322& CCD& (26)&   445& 55999.0727& 162020& 0.003606& CCD& (26)\\
421& 55621.1986& 157627& 0.002604& CCD& (26)&   446& 55999.1547& 162021& -0.000411& CCD& (26)\\
422& 55937.3981& 161303& 0.003264& CCD& (26)&   447& 55999.2375& 162022& -0.003628& CCD& (26)\\
423& 55961.3051& 161581& -0.002489& CCD& (26)&  448& 56000.0161& 162031& 0.000818& CCD& (26)\\
424& 55961.3907& 161582& -0.002906& CCD& (26)&  449& 56000.0978& 162032& -0.003499& CCD& (26)\\
425& 55964.3148& 161616& -0.003387& CCD& (26)&  450& 56000.1874& 162033& 0.000084& CCD& (26)\\
451& 56000.2762& 162034& 0.002867& CCD& (26)&   457& 56063.0691& 162764& 0.003287& CCD& (26)\\
452& 56001.0459& 162043& -0.001587& CCD& (26)&  458& 56064.0137& 162775& 0.001699& CCD& (26)\\
453& 56001.1368& 162044& 0.003296& CCD& (26)&   459& 56064.0948& 162776& -0.003218& CCD& (26)\\
454& 56001.2186& 162045& -0.000921& CCD& (26)&  460& 56065.0419& 162787& -0.002306& CCD& (26)\\
455& 56001.3023& 162046& -0.003238& CCD& (26)&  461& 56068.0580& 162822& 0.003196& CCD& (26)\\
456& 56062.0312& 162752& -0.002407& CCD& (26)&  & & & & & \\

\enddata
\tablecomments{
Source: (1) \citet{Tsesevich1973}; (2) \citet{Filatov1960}; (3)
\citet{Pocs2001}; (4) \citet{Broglia1975}; (5) \citet{Braune1979}; (6)
\citet{Braune1982}; (7) \citet{Huebscher1985}; (8)
\citet{Rodriguez1992}; (9) \citet{Huebscher1992}; (10)
\citet{Hintz1997}; (11) \citet{Agerer1999}; (12) \citet{199905IBVS};
(13) \citet{2001IBVS}; (14) \citet{Zhou2001}; (15) \citet{2003IBVS};
(16)\citet{200508IBVS}; (17) \citet{200510IBVS}; (18)
\citet{200605IBVS}; (19) \citet{200611IBVS}; (20) \citet{200703IBVS};
(21) \citet{2010JAVSO}; (22) \citet{200710IBVS}; (23)
\citet{2009IBVS}; (24) \citet{2010IBVS}; (25) \citet{2011BAVSM}; (26) this work.}
\label{tab7}
\end{deluxetable}

In addition, we made a quadratic fit with a second-order polynomial,
\begin{equation} \begin{split}
  HJD_{max} &= 2442062.5822(\pm 0.0002)\\
   &+ 0.086017060(\pm 0.000000006) E \\
   & + 0.5 \times 1.09(\pm 0.38) \times 10^{-13}E^{2}
\end{split} \end{equation}
 with the standard deviation of $\sigma_{1} = 0.00244$ days. The quadratic terms differ from zero by a factor of 2.87$\sigma$ with the significance of $\sim 99.5$ percent and the statistic test proposed by \citep{Pringle1975} suggests that the small improvement in period deviation gives the quadratic term in the fit with a significance of $\sim 99.3$ percent. From the values in equation~(4), we take the period change rate of AE UMa as $(1/P_{0})(dP_{0}/dt) = 5.4(\pm 1.9) \times 10^{-9}$ yr$^{-1}$ that is different from the result of $-0.35 \times 10^{-10}$ yr$^{-1}$ provided by \citet{Zhou2001}.  This value will be used in our model calculations in the next section. However, the data may not be distributed as Gaussian random noise and more data points need to be collected to confirm this period change.

Since the modulation frequency $f_{m} = f_{1}-f_{0}$ has not been varying  significantly comparing to $f_{1}$ \citep{Pocs2001},  one may take the method which used in \citet{Pocs2001} and \citet{Zhou2001} to calculate the rate of changes of the first overtone frequency. But the result show large uncertainties. Hence, we do not consider it a credible result from our observations.

\section{CONSTRAINTS FROM THE THEORETICAL MODELS}

Because the order of the period change of the fundamental mode is the same as the result from our calculation in Section 4 in the post-MS phase (about $10^{-8}\ yr^{-1}$), then we assume that the result is completely from the evolutionary effects.

In this section, we describe the details of calculation of the theoretical models of AE UMa  to constrain the physical parameters for the target. Subsection 5.1 presents the initial input physical parameters of AE UMa for the theoretical models, Subsection 5.2 uses the two frequencies $f_{0}$ and $f_{1}$ to constrain the initial parameters and determines some parameters for the subsequent calculation; Subsection 5.3 uses two independent ways to calculate the period changes of the fundamental mode of AE UMa induced by the stellar evolutionary effects.

\subsection{Physical Parameters}

\citet{Rodriguez1992} made $uvby\beta$ photoelectric photometry for AE UMa. Intrinsic values of $b - y$, $m_{1}$ and $c_{1}$ were derived and the stellar physical parameters were determined. The effective temperature of AE UMa varied from 8320 K to 7150 K. The surface gravity $\log g$ varied from $4.16$ to $3.77$. The mean values obtained along the cycle were $<T_{\rm eff}> = 7560$ K and $<\log g> = 3.90$, respectively. The metal abundance was estimated from $\delta m_{1}$ at the minimum light as $[Fe/H] = -0.3$.
By using the $\log g - \log P$ relation derived by \citet{Claret1990}, \citet{Rodriguez1992} obtained the values of $M = 1.80\ M_{\odot}$, $Age = 1.3 \times 10^{9}$ yr and $M_{bol} = 1^{m}.76$.
\citet{Hintz1997} provided the value of $[Fe/H] = -0.1$ according to the relation between the $P_{1}/P_{0}$ ratio and the $[Fe/H]$ value for dwarf Cepheids which was derived from \citet{Hintz19971}. They got the $[Fe/H]$ values ranging from $-0.4$ to $-0.1$.
We listed the parameters of AE UMa from \citet{Rodriguez1992} and \citet{Hintz1997} in Table \ref{tab8}.  The atmospheric parameters derived from our spectrum are in good agreement with that above values, in particular for the metal ratio $[Fe/H]$ (comparison of Table~\ref{tab3} to \ref{tab11}). We note that, in order to perform a search for the best fitting model in a wide parametric range, we used $3\sigma$ as the intervals of constraints for our theoretical calculation as follows.

\begin{table*}
 \caption{Physical parameters of AE UMa from \citet{Rodriguez1992} and \citet{Hintz1997}. The $\log(L/L_{\odot})$ value was derived based on $\log(L/L_{\odot}) = 0.4 (M_{bol_{\odot}} - M_{bol})$.}
  \centering

 \begin{tabular}{c|c|c|c}
 \hline
  Parameters  & Mean value & Intervals & $3 \sigma$ \\

 \hline
 $[Fe/H]$ & -0.3  & [-0.4,-0.1] & ---    \\
 $T_\mathrm{eff}$ (K) & 7569  & [7150,8320] & [5980,9490]   \\
 $\log g$  & 3.90  & [3.77,4.16] & [3.38,4.55]    \\
 $M_{bol}$ & ---   & [1.53,1.93]  & [1.33,2.13]  \\
 $M(M_{\odot})$      & ---   & [1.75,1.95] & ---    \\
 $\log(L/L_{\odot})$ & --- & [1.16,1.32] & [1.08,1.40] \\
 \hline
 \end{tabular}
\label{tab8}
\end{table*}

\subsection{Constraints from $f_{0}$ and $f_{1}$}

Modules for Experiments in Stellar Astrophysics (MESA) is a suite of source-open, robust, efficient, thread-safe libraries for a wide range of applications in computational stellar astrophysics \citep{mesa2011,mesa2013}. The 1-D stellar evolution module, MESA star, combines many of the numerical and physics modules for simulations of a wide range of stellar evolution scenarios ranging from very-low mass to massive stars, including advanced evolutionary phases. The ''\textit{astero}'' extension to MESA star implements an integrated approach that passes results automatically between MESA star and the new MESA module based on the adiabatic code ADIPLS \citep{Christensen2008}.

In MESA version 6208, the \textit{astero} extension enables calculation of selected pulsation frequencies by MESA star during the evolution of the model. This allows fitting to the observations that can include spectroscopic constraints (e.g., $[Fe/H]$, $\log g$ and $T_{eff}$ ), asteroseismic constraints, the large frequency separation ($\Delta\nu$) and the frequency of maximum power ($\nu_{max}$), and even individual frequencies observed. For the automated $\chi^{2}$ minimization, \textit{astero} will evolve a pre-main sequence model from a user defined starting point, and find the best match along that single evolutionary track. The code then recalculates the track, again initiated at the pre-main sequence, with different initial parameters such as mass, chemical composition, mixing length parameter and overshooting, and repeats until the minimum $\chi^{2}$ is found.

We used the scan-grid mode to minimize the $\chi^{2}$ for each model, which helps to compact the intervals of the physical parameters.

Every model of evolution starts with creating a pre-main-sequence model by specifying the mass, $M$,  at a uniform composition. The equation-of-state tables are constructed from the 2005 update of the OPAL EOS tables \citep{Rogers2002} and SCVH tables \citep{Saumon1995}. The MESA opacity tables, which are derived from Type 1 and 2 OPAL tables \citep{Iglesias1993,Iglesias1996}, tables from OP \citet{Seaton2005}, and \citet{Ferguson2005}, cover a large range $2.7 \leqslant \log T \leqslant 10.3$ and $-8 \leqslant \log R \leqslant 8$. The hydrogen burning reaction rates in the calculations are from \citet{Bahcall1997, Bahcall2002}.

MESA star treats convective mixing as a time-dependent, diffusive process with a diffusion coefficient D as
\begin{equation}
D_{OV} = D_{conv,0} \exp(-\frac{2z}{f \lambda_{P,0}}),
\end{equation}
where $\lambda_{P,0}$ is the pressure scale height at that location, $z$ is the distance in the radiative layer away from that location, and f is an adjustable parameter \citep{Herwig2000}.

In all our subsequent calculation, the used opacity and EOS tables are $eos\_file\_prefix = mesa$, $kappa\_file\_prefix = gs98$ and $kappa\_lowT\_prefix = lowT\_Freedman11$. The atmosphere model is $which\_atm\_option = photosphere\_tables$ photosphere.

The mixing-length parameter $\alpha_{MLT}$ was chosen as 1.89, since the choice has  actually a very small effect on our models \citep{Yang2012}. The convective overshooting parameter $f_{ov} = 0.015$ was the initial value of MESA (version 6208) . The effects of rotation on the evolutionary period changes are disregarded, concerning AE UMa as a HADS which are very slow rotators \citep{Breger2000review}. Table \ref{tab9} lists the parameters of the grid of model to search for $f_{0}$ and $f_{1}$ of AE UMa. The diffusion effects were not taken into account because of its negligible results on the models with mass $1.30\ M_{\odot}$ to $2.70\ M_{\odot}$ after the main sequence and the post-main sequence (before Red Giant phase).

\begin{table*}
\caption{The parameters of the grid of model to search for $f_{0}$ and $f_{1}$. Since the values of $M_{bol}$ and $\log(L/L_{\odot})$ in Table \ref{tab6} were calculated from the stellar models \citet{Rodriguez1992} other than from observations, which depended on the models they used, we did not use these values as the constraints during our calculation. }
\centering 

 \begin{tabular}{c|c|c|c}
 \hline
 Parameters & Maximum & Minimum & Step  \\
 \hline
 $[Fe/H]$  & -0.1 & -0.4 & 0.05 \\
 initial Y  & 0.33 & 0.23 & 0.02 \\
 Mass  & 2.7 & 1.3 & 0.02 \\
$\log T_\mathrm{eff}$ & 3.977 & 3.777 & ---\\
$\log g$ & 4.55 & 3.38  & ---\\
 \hline
 \end{tabular}
\label{tab9}
 \end{table*}

As the result, we got the models which included the frequencies $f_{0}$ and $f_{1}$ along with the stellar evolution tracks. These tracks provided the relevant intervals of the parameters for subsequent calculation, as listed in Table \ref{tab10}.

\begin{table*}
\caption{The parameters determined with the constraints from $f_{0}$ and $f_{1}$. The grid was constructed also within the parameters intervals of $T_{eff}$ and $\log g$ listed in Table \ref{tab6}. In order to show an obvious comparison of the evolutionary tracks, we calculated the tracks with the stellar mass from $1.30$ to $2.30 M_{\odot}$.}
  \centering
 \begin{tabular}{c|c|c}
 \hline
 Parameters & Maximum & Minimum  \\
 \hline
 $[Fe/H]$  & -0.2 & -0.4  \\
 initial Y  & 0.27 & 0.23  \\
 Mass  & 2.26 & 1.32  \\
 \hline
 \end{tabular}
\label{tab10}
\end{table*}

With the observations and the method used, one can believe that $[Fe/H]$ and $\log g$ values are in good reliability \citep{Stromgren1956,Crawford1966}. So, we decide to take the value $[Fe/H] = -0.3$ in our calculation.

We used the formula (with the solar metallicity  $X_{\odot} = 0.7381$, $Y_{\odot} = 0.2485$ and $Z_{\odot} = 0.0134$ from \citet{Asplund2009}),
\begin{equation}
  [Fe/H] = \log(\frac{Z}{X}) - \log(\frac{Z_{\odot}}{X_{\odot}})
\end{equation}
and the formula,
\begin{equation}
  X + Y + Z = 1
\end{equation}
to calculate the initial Z. In \citet{Girardi2000}, a model was calculated with a couple of value $(Y,Z) = (0.25,0.008)$, which accords with the values in our previous calculation (derived with MESA \textit{astero} with the value of $[Fe/H]$).

At last, by integrating all the information about the value of $(Y,Z)$, we decide to choose $(Y,Z) =  (0.25,0.008521)$ as the unique initial value for the subsequent calculation.

\subsection{Constraints from the Period Variation}

Not like the solar-like stars for which many frequencies are detected,  most HADS are observed with only the fundamental and the first overtone modes in general \citep[e.g.,][]{balona2012,Ulusoy2013}. As a result, the period variation becomes a very important constraint on the model calculation of these stars. Detection of high precision of period variation may offer strong constraints on AE UMa. We used two independent ways to calculate the period variations of AE UMa theoretically.

\subsubsection{Calculation from Stellar Evolutionary Effect}

The variation rate of the fundamental period derived from long time-scale of observations of AE UMa shows a positive period change. From the theoretical point of view, the period changes caused by stellar evolution in and across the lower instability strip permit an observational test of stellar evolution theory \citep{Breger1998}.

The period-luminosity-color relation can be expressed as \citep{Breger1998}
\begin{equation}
\log P = -0.3 M_{bol} - 3 \log T_{eff} - 0.5 \log M + \log Q + constant,
\end{equation}
where $P$ is the period of a radial mode of pulsation, $M_{bol}$ is the bolometric absolute magnitude, $T_{eff}$ is the effective temperature, $M$ is the stellar mass in solar mass, and $Q$ is the pulsation constant in days. For $\delta$ Scuti stars with radial pulsation, the constant is 12.708. For individual stars, evolutionary period changes over long time scales. An evolutionary change in $T_{\rm eff}$, $M_{bol}$, and $M$ leads to a period change of

\begin{equation}
\frac{1}{P} \frac{dP}{dt} = -0.69\frac{dM_{bol}}{dt} - \frac{3}{T_{\rm eff}} - 0.5 \frac{1}{M} \frac{dM}{dt} + \frac{1}{Q}\frac{dQ}{dt}.
\end{equation}

Assuming that the stellar mass $M = constant$ for $\delta$ Scuti stars during the observation interval with mass of $1.30-2.30\ M_{\odot}$ and that the variation of pulsation constant is negligible as a very small quantity, \citet{Yang2012} got,
\begin{equation}
  \frac{1}{P} \frac{dP}{dt} \approx -0.69\frac{dM_{bol}}{dt} - \frac{3}{T_{\rm eff}} \frac{dT_{\rm eff}}{dt}.
\end{equation}

As indicated by \citet{Rodriguez2001}, the HADS locate on or near the main sequence of the H-R diagram. Consequently, the evolutionary models are constructed from the pre-main-sequence Hayashi phase to the end of the  main sequence. The effect of rotation was not considered here for mainly two reasons: (1) the HADS stars are typically with slow rotation and most have a projected rotational velocity, $V\sin i$,  around 20~km\,s$^{-1}$ \citep[see, e.g.,][]{Solano1997}; (2) the effects of rotation with speed of $V\sin i = 18$~km\,s$^{-1}$ in HADS is very similar to that of absence of rotation \citep{Casas2006}.

\begin{figure*}
\centering
\includegraphics[width=0.8\hsize,height=0.5\hsize]{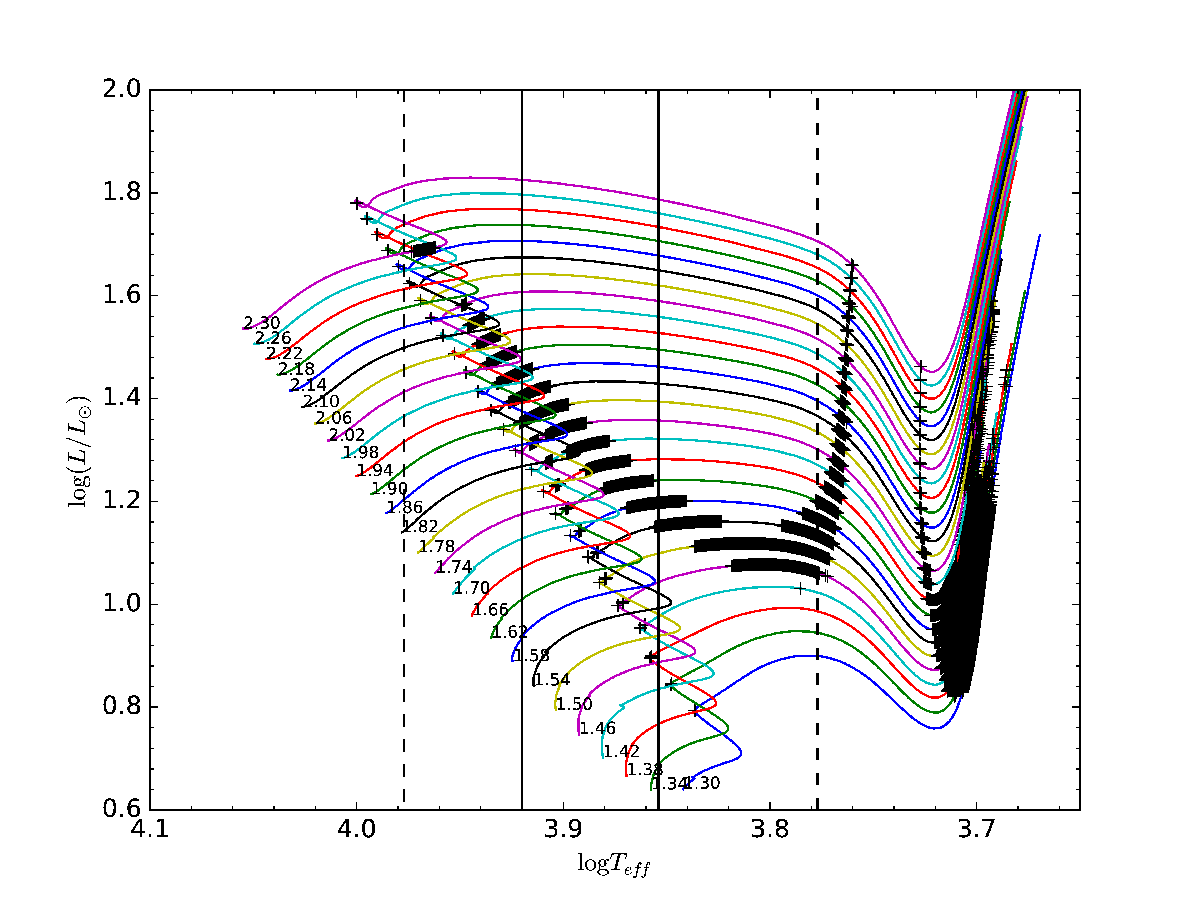}
\caption{Evolutionary tracks of models with mass from $1.30\ M_{\odot}$ to $2.30\ M_{\odot}$ for $(Y,Z) = (0.25,0.008521)$. The solid and dashed vertical lines on the H-R diagram are determined from the observed $T_{eff}$ in $1 \sigma$ and $3\sigma$, respectively. The Marks on the tracks indicate the models with the values of  the evolutionary period changes of $(1/P_{0})(dP_{0}/dt)$  inside the interval $[3.5210\times10^{-9}, 7.2448\times10^{-9}]$ in units of $yr^{-1}$. Please note that, the tracks are shown on the diagram with the mass interval of $0.04 M_{\odot}$.}
\label{fig7}
\end{figure*}

The evolutionary tracks we constructed from $1.30\ M_{\odot}$ to $2.30\ M_{\odot}$ are shown in Figure \ref{fig7}, and the corresponding variation rates of the period are marked.

As shown in Figure \ref{fig7}, the states whose values of period changes are consistent with the observed ones determined from the $O-C$ analysis lie just after the second turn-offs leaving the main sequences on the evolutionary tracks.

\subsubsection{Calculation from ADIPLS}

ADIPLS (the Aarhus adiabatic oscillation package) is a programe for calculation of adiabatic oscillations of stellar models \citep{Christensen2008}. We used it to calculate the frequencies of the eigen modes of the model at each step of our evolutionary state. As a result, we got the frequencies of the model of $F_{0}$ and $F_{1}$ then deduced the variation of the frequencies of $F_{0}$ and $F_{1}$. In the calculation, the input frequencies $F_{0}$ and $F_{1}$ are the fundamental and first overtone frequencies with quantum numbers of $l=0$ and $n=1,2$, respectively.

\begin{figure*}
\centering
\subfigure[]{
           \centering
           \includegraphics[width=0.6\textwidth,height=0.25\textheight]{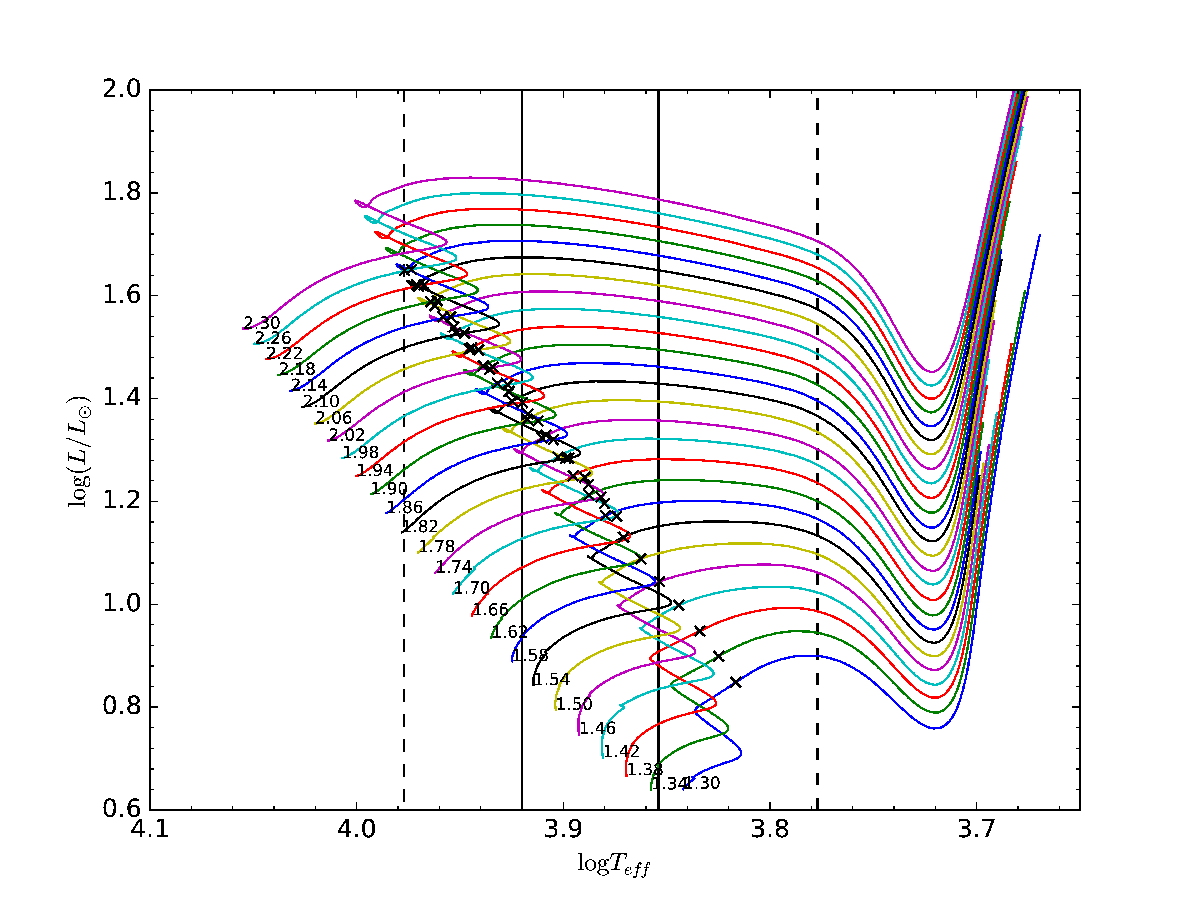}
            }
            \centering
\subfigure[]{
             \centering
             \includegraphics[width=0.6\textwidth,height=0.25\textheight]{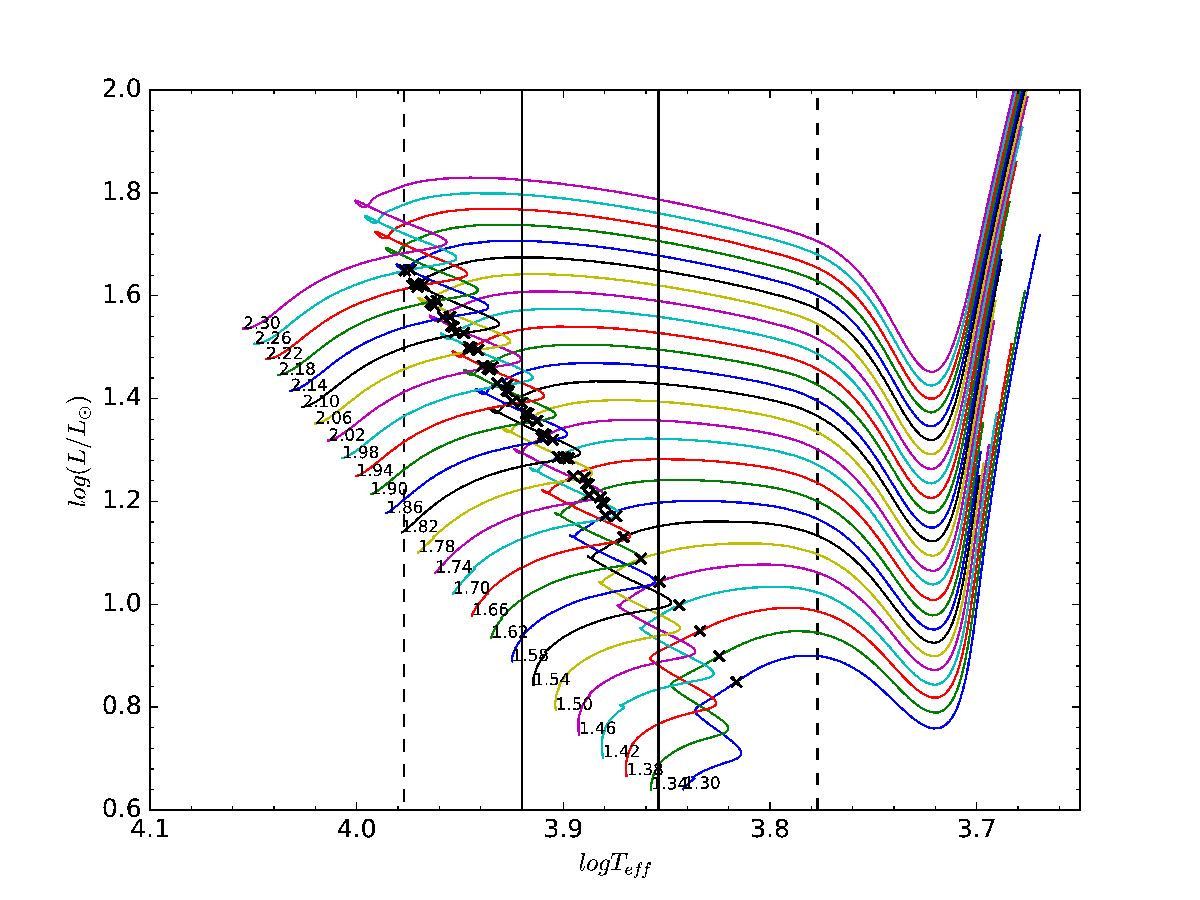}
            }
\subfigure[]{
           \centering
           \includegraphics[width=0.6\textwidth,height=0.25\textheight]{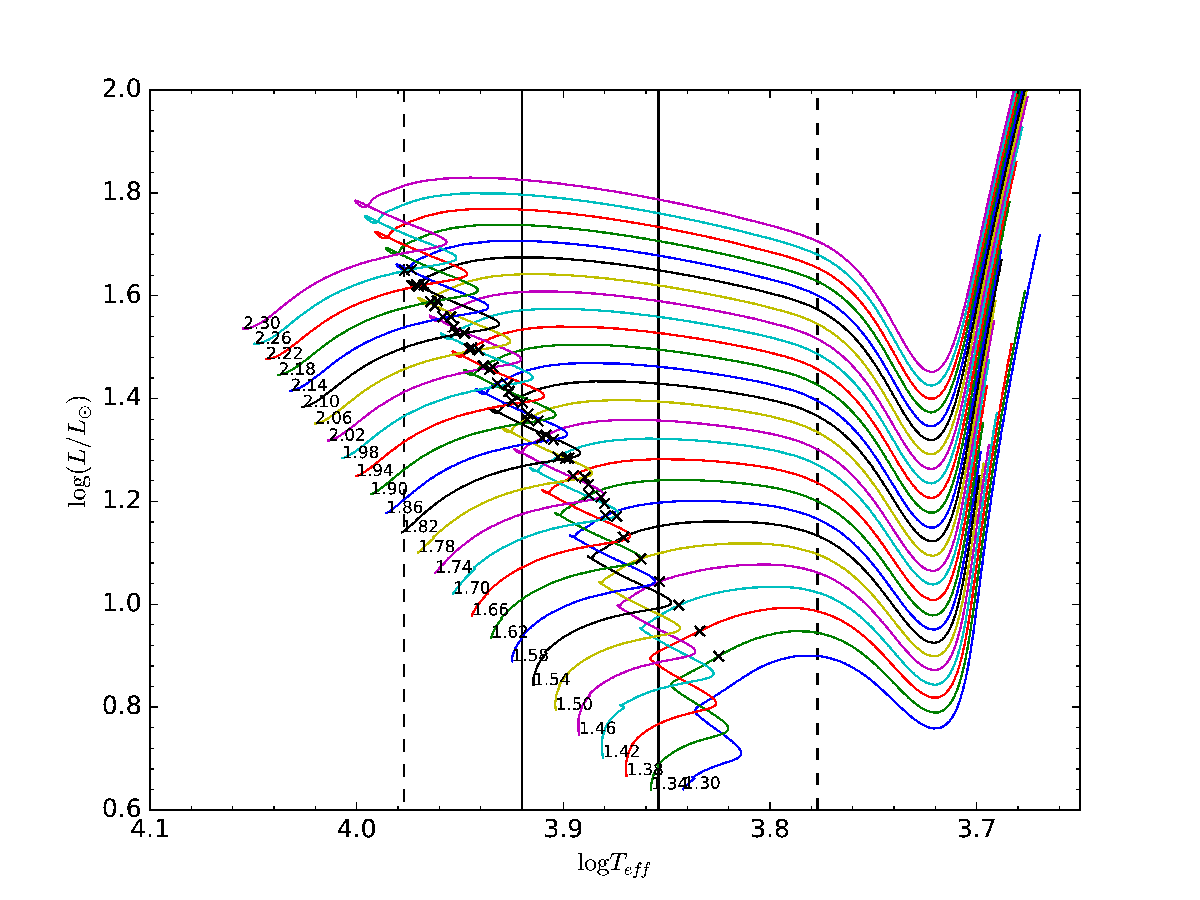}
           }
\caption{Models with the values of $F_{0} $ and $F_{1}$ calculated from ADIPLS consistent with the observed ones ($F_{0} \in [f_{0} - 3\sigma,f_{0} + 3\sigma$] and $F_{1} \in [f_{1} - 3\sigma,f_{1} + 3\sigma$]) are marked on the evolutionary tracks. (a): for $F_{0} \in [f_{0} - 3\sigma,f_{0} + 3\sigma$]; (b): for $F_{1} \in [f_{1} - 3\sigma,f_{1} + 3\sigma$]; (c): for both  $F_{0} \in [f_{0} - 3\sigma,f_{0} + 3\sigma]$ and $F_{1} \in [f_{1} - 3\sigma,f_{1} + 3\sigma]$.}
\label{fig8}
\end{figure*}

We calculated the evolutionary tracks from $1.30\ M_{\odot}$ to $2.30\ M_{\odot}$, and got the frequency values on each tracks, as shown in Figure \ref{fig8}. Figure~\ref{fig8} (a) and (b) show that the models with appropriate frequencies for AE UMa could appear (i) just before the first turn-offs, (ii) after the first and before the second turn-offs, (iii) just after the second turn-offs. We integrated the constraints from $f_{0}$ and $f_{1}$ and got the results shown in Figure \ref{fig8} (c).

\begin{figure*}
\centering
           \centering
           \includegraphics[width=0.8\textwidth,height=0.4\textheight]{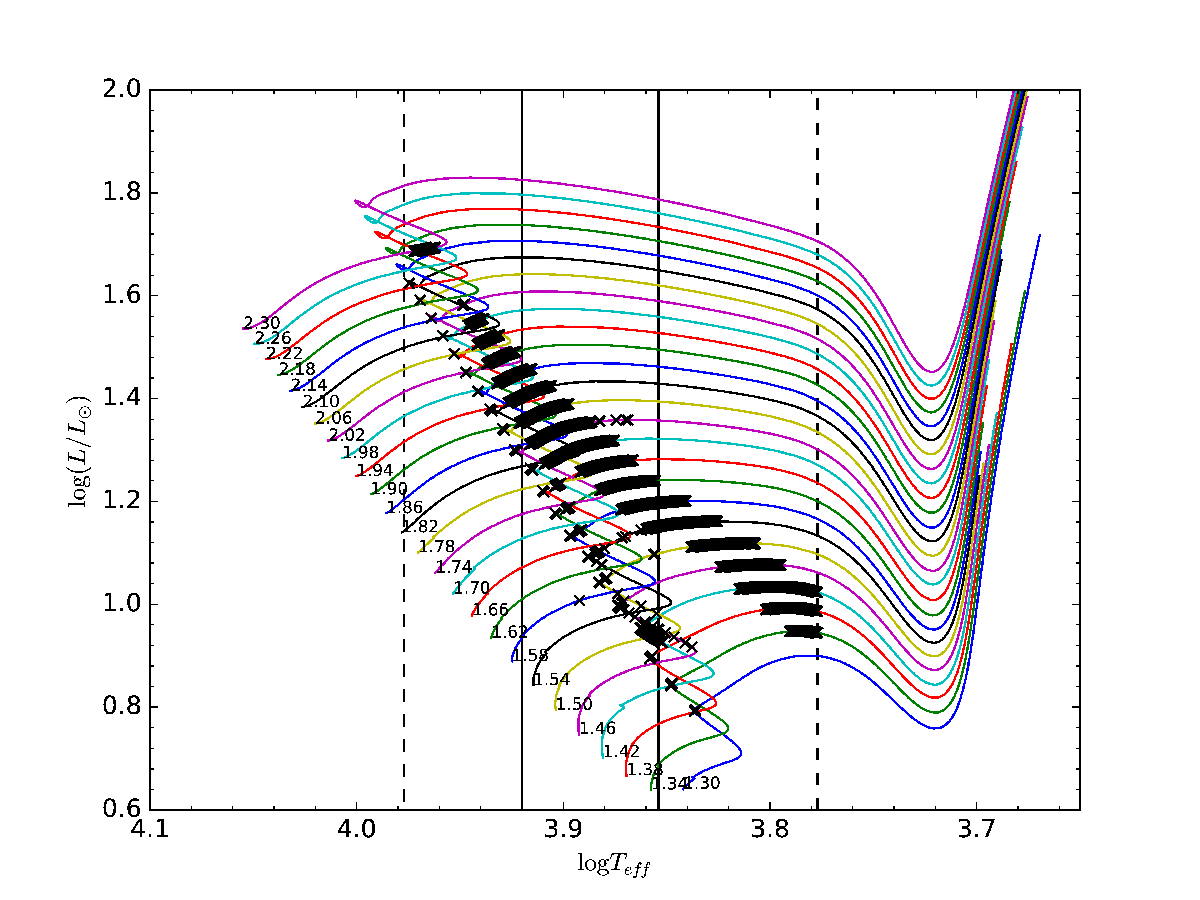}

\caption{The models for which the period variations of the fundamental mode $(1/P_{0})(dP_{0}/dt)$ calculated with ADIPLS agree with the observed values of $5.3829(\pm 1.8619) \times 10^{-9}\ yr^{-1}$ are marked on the evolutionary tracks. }
\label{fig9}
\end{figure*}

In addition, we also calculated the variations of the frequencies of the eigen-modes of the models by using ADIPLS. Adding the constraint from $(1/P_{0})(dP_{0}/dt)$ which are thought to be due to the evolutionary effects, we got the results in Figure \ref{fig9}.

One can find that this result is almost consistent with the result from Figure \ref{fig7} just after the second turn-off. The differences come from we did not consider the variation of the stellar mass and the pulsation constant along the evolutionary tracks in Figure \ref{fig7}.

\begin{figure*}
\subfigure[]{
           \centering
           \includegraphics[width=0.8\textwidth,height=0.4\textheight]{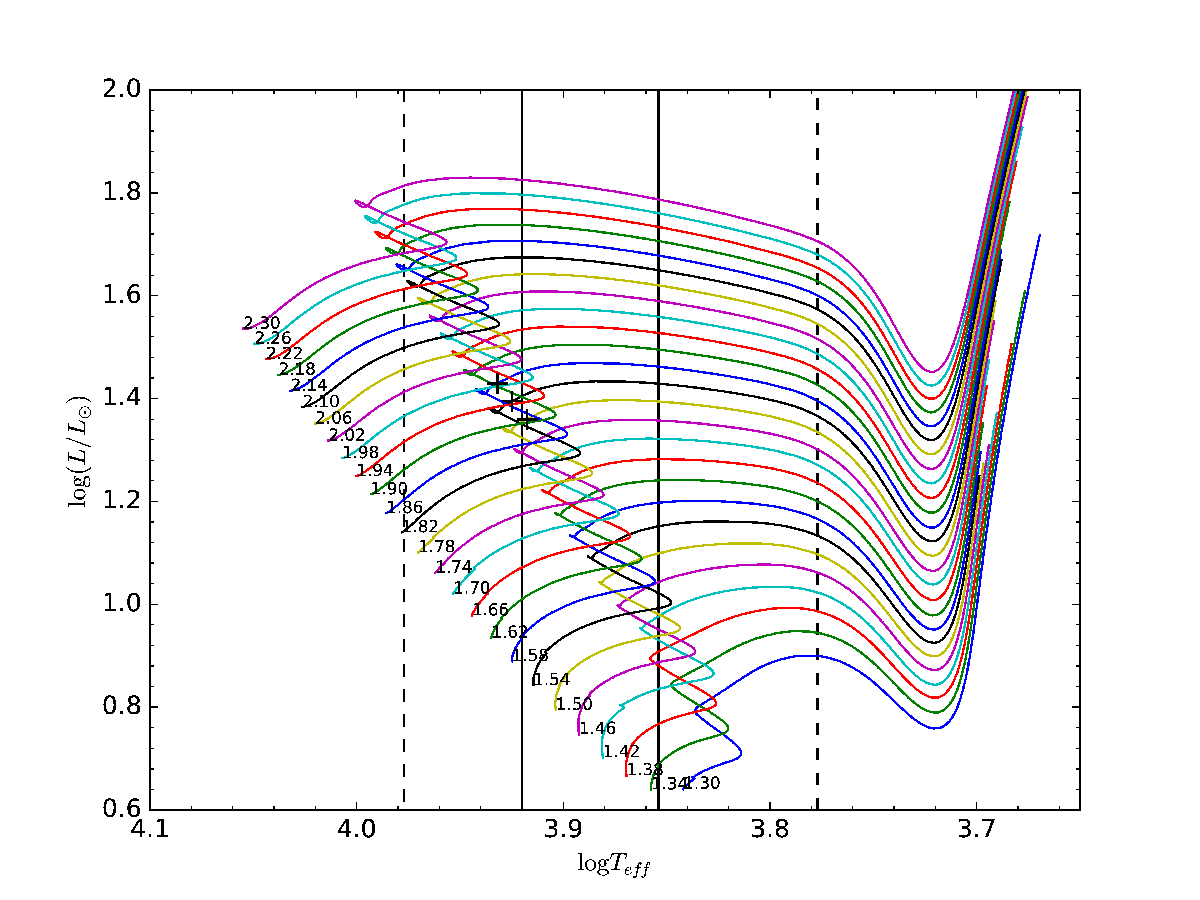}
}
\subfigure[]{
           \centering
           \includegraphics[width=0.8\textwidth,height=0.4\textheight]{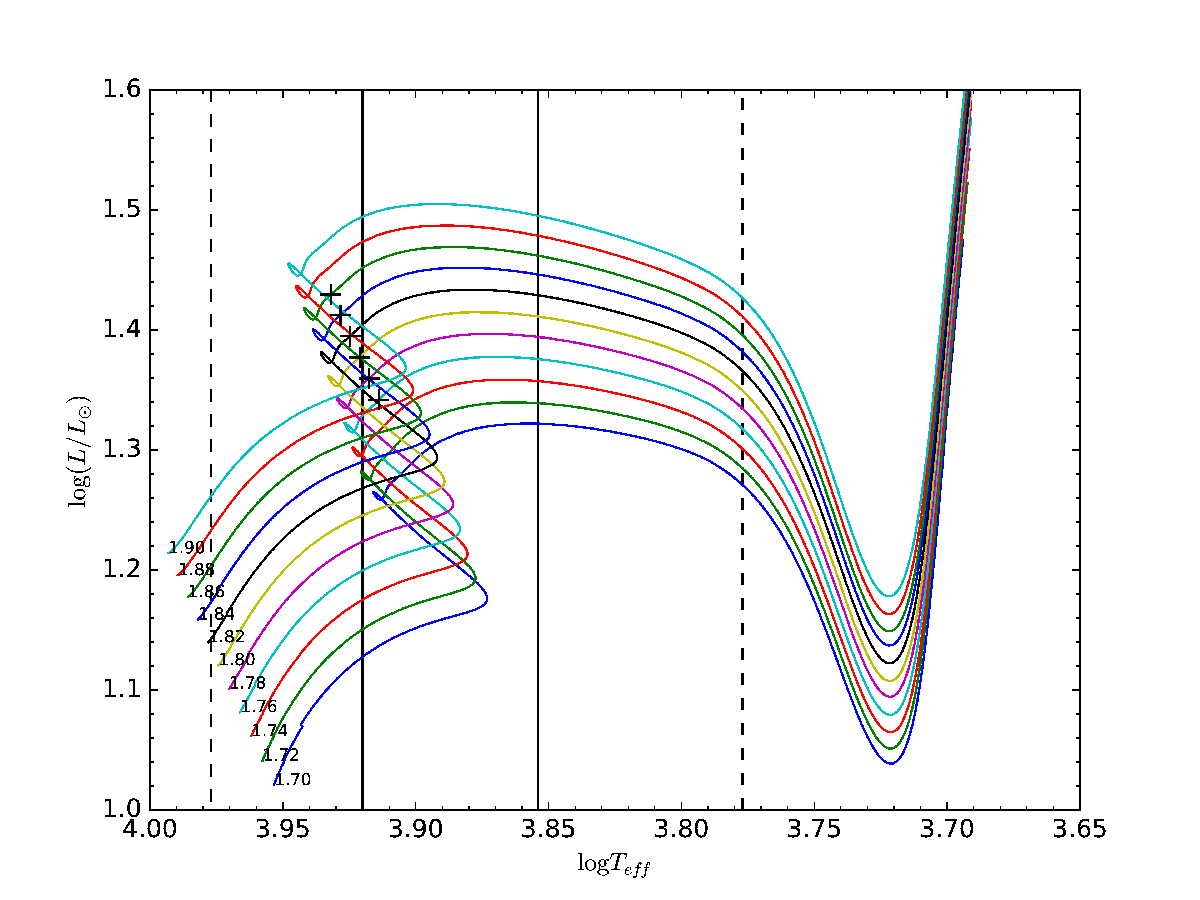}
            }

\caption{Evolutionary tracks of star models. (a):  marks on the tracks indicate the models with constraints from $f_{0}$, $f_{1}$ and $(1/P_{0})(dP_{0}/dt)$. (b): A zoom in of the (a).}
\label{fig10}
\end{figure*}

Finally, we combined the constraints from $f_{0}$, $f_{1}$ and $(1/P_{0})(dP_{0}/dt)$, and got the results which is shown in Figure \ref{fig10}.

One concludes that AE UMa should locate after the second turn-offs of the evolutionary tracks leaving the MS. Hence, one finds that the period variations of the fundamental mode of AE UMa are caused by the evolutionary effect. The rate of variation is consistent with the theoretically predicted value by \citet{Breger1998}.

With the discussion above and constraints from the physical parameters, one can conclude that the mass of AE UMa ranges from $1.75\ M_{\odot}$ to $1.86\ M_{\odot}$, the age from $0.96 \times 10^{9}$ yr to $1.15 \times 10^{9}$ yr.

\begin{figure*}
\centering
           \centering
           \includegraphics[width=0.8\textwidth,height=0.5\textheight]{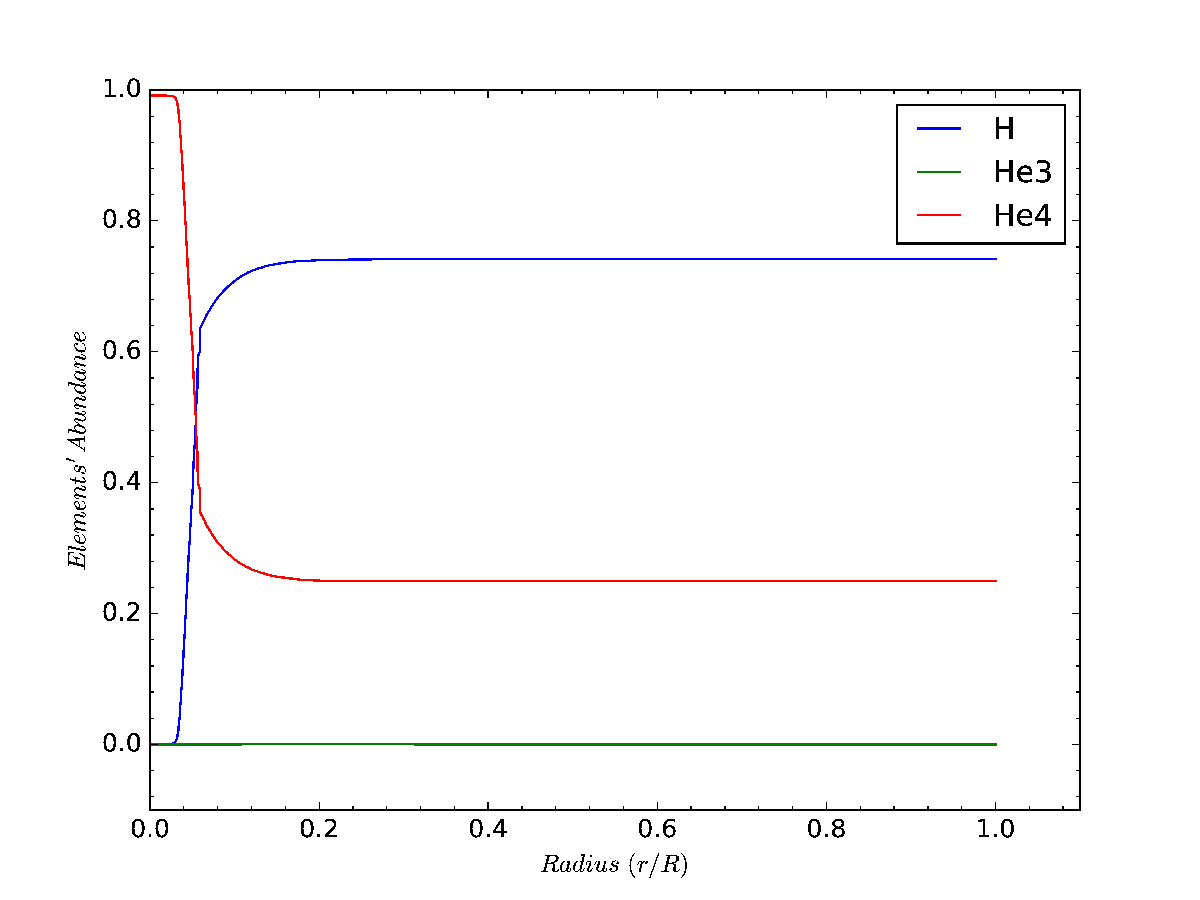}
\caption{Elements' Abundance distribution of $H$, $He^{3}$ and $He^{4}$ inside the star for the model with the star mass of  $1.80M_{\odot}$.}
\label{fig11}
\end{figure*}

\begin{figure*}
\centering
           \centering
           \includegraphics[width=0.8\textwidth,height=0.5\textheight]{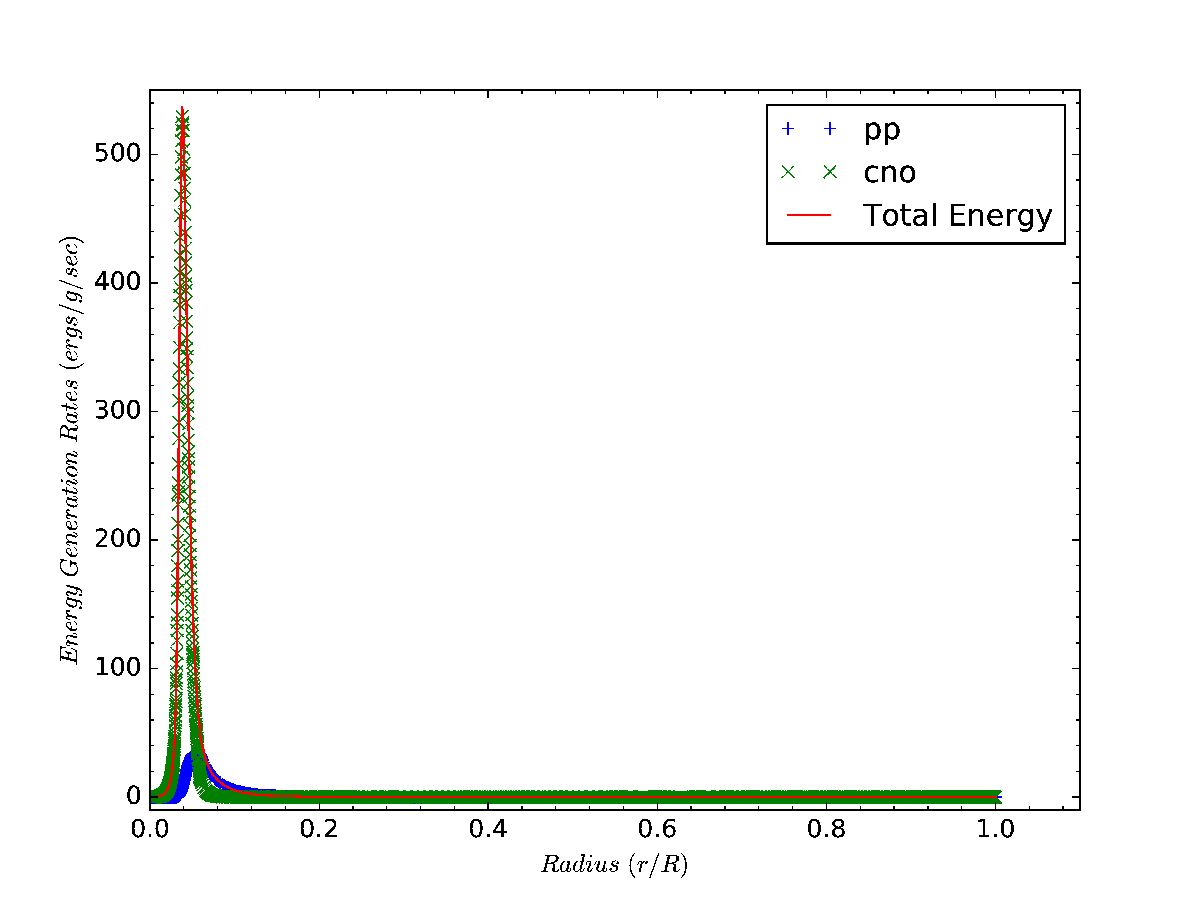}
\caption{Energy generation rates distribution inside the star for the model with the star mass of $1.80M_{\odot}$.}
\label{fig12}
\end{figure*}

We chose $1.80\ M_{\odot}$ as a sample to study the evolutionary state and the interior of the models that we gained. More details of the parameters we got from calculation are listed in Table \ref{tab11}. Figure \ref{fig11} shows the distribution of $H$, $He^{3}$ and $He^{4}$ versus the stellar radius. Figure \ref{fig12} presents the energy distribution inside the star. From Figure \ref{fig11} and \ref{fig12}, one may find that the star should have a helium core and a hydrogen-burning shell.

\begin{table*}
\caption{Physical parameters of AE UMa obtained from our calculation.}
\centering

\begin{tabular}{c|c|c}
\hline
Parameter    & Value    & Uncertainty ( \% ) \\
\hline
Mass ($M_{\odot}$)&  $1.805 \pm 0.055$       &   3.04              \\
Age ($10^{9}$ yr)&    $1.055 \pm 0.095$     &        9.00         \\
$\log T_\mathrm{eff}$ &   $3.922 \pm 0.01$      &      0.25           \\
Radius ($10^{11}$~cm)&  $1.647 \pm 0.032 $       &    1.94             \\
$\log g$ &  $3.9543 \pm 0.0044$   &      0.11           \\
$\log L$ &  $1.381 \pm 0.048$       &      3.51           \\
\hline
\end{tabular}
\label{tab11}
\end{table*}

We also calculated models with different values of $Y$, $Z$ and $\alpha_{ov}$ by using different grids. The result showed that the states were not different significantly. All the results pointed out that AE UMa should lie just after the second turn-off with a helium core and a hydrogen-burning shell.


\section{CONCLUSIONS}

We analyse the photometric data gathered on AE~UMa with 40 nights spanning over from 2009 to 2012 and detect 37 frequencies above the so-called 4$\sigma$ detection threshold, among which 25 frequencies are newly detected.
All these frequencies are linked to be either harmonics or linear combinations of the two main frequencies $f_{0} = 11.62560\ c\ d^{-1}$ and $f_{1} = 15.03123\ c\ d^{-1}$, corresponding to the fundamental and the first overtone radial pulsation modes, respectively. No frequencies of the other pulsation modes were detected from the observed data.

An $O-C$ diagram is constructed with combination of the 84 times of maximum light determined from our new observations and 360 ones listed in the literature, leading to the updated value of period $P_{0} = 0.0860170781$ days. A new ephemeris with a quadratic solution suggests that the  period change rate of the fundamental mode of AE UMa is of $(1/P_{0})(dP_{0}/dt) = 5.4(\pm 1.9) \times 10^{-9}$ yr$^{-1}$. The value is different with the result obtained by \citet{Zhou2001} and need to be confirmed with more data that will be collected from observations in the near future, both from ground and space \citep[e.g., TESS,][]{2014ricker}. Because the large values of the derivative of $(1/P_{1})(dP_{1}/dt)$ obtained from the standard $O-C$ method, we did not use this value as a constraint in the model calculation.

With the spectroscopic observation data, we got the low-resolution spectrum and used the automated 1D parametrization pipeline LASP to obtain the stellar atmospheric parameters of AE UMa. These parameters (especially the $[Fe/H]$ value of -0.32) certificate that AE UMa is a Pop. I $\delta$ Scuti star rather than a Pop. II SX Phe star.

We then calculated models of stars with masses between $1.30\ M_{\odot}$ and $2.70\ M_{\odot}$. With the constraints of the values of $f_{0}$, $f_{1}$ and $(1/P_{0})(dP_{0}/dt)$, we conclude that AE UMa lies just after the second turn-offs of the evolutionary tracks leaving the main sequence. The corresponding mass should be $1.805 \pm 0.055\ M_{\odot}$ and the age $1.055 \pm 0.095 \times 10^{9}$ yr. At this evolutionary phase, the star should have a helium core and a hydrogen-burning shell.

Moreover, according to the concrete observational evidence, we provide an example of the HADS whose evolutionary stage is on the post-main-sequence. This gives a direct support to the general consensus that $\delta$ Scuti stars are probably normal stars evolving in the main-sequence or the immediate post-main-sequence stages.

\section*{Acknowledgments}

JSN and JNF acknowledge the support from the Joint Fund of Astronomy of National Natural Science Foundation of China (NSFC) and Chinese Academy of Sciences through the grant U1231202, the NSFC grant 11673003, the National Basic Research Program of China (973 Program 2014CB845700 and 2013CB834900), and the LAMOST FELLOWSHIP supported by Special Funding for Advanced Users, budgeted and administrated by Center for Astronomical Mega-Science, Chinese Academy of Sciences (CAMS). JSN thanks Nami Mowlavi for helpful discussions and useful advice.


\end{document}